\documentclass[fleqn,usenatbib]{mnras}

\usepackage{newtxtext,newtxmath}

\usepackage[T1]{fontenc}
\usepackage{ae,aecompl}

\usepackage{graphicx}	
\usepackage{amsmath}	
\usepackage{amssymb}	

\title[Fast cloud collisions in a strong bar]{Fast cloud-cloud collisions in a strongly barred galaxy: Suppression of massive star formation}

\author[Y. Fujimoto et al.]{
Yusuke Fujimoto,$^{1,2}$\thanks{E-mail: yfujimoto@carnegiescience.edu}
Fumiya Maeda,$^{3}$
Asao Habe$^{4}$
and Kouji Ohta$^{3}$
\\
$^{1}$Research School of Astronomy \& Astrophysics, Australian National University, Canberra, Australian Capital Territory 2611, Australia\\
$^{2}$Earth and Planets Laboratory, Carnegie Institution for Science, 5241 Broad Branch Road, NW, Washington, DC 20015, USA\\
$^{3}$Department of Astronomy, Kyoto University, Kitashirakawa-Oiwake-Cho, Sakyo-ku, Kyoto, Kyoto 606-8502, Japan\\
$^{4}$Graduate School of Science, Hokkaido University, Kita 10 Nishi 8, Kita-ku, Sapporo, Hokkaido 060-0810, Japan
}

\date{Accepted XXX. Received YYY; in original form ZZZ}

\pubyear{2019}

\begin{document}
\label{firstpage}
\pagerange{\pageref{firstpage}--\pageref{lastpage}}
\maketitle

\begin{abstract}
    Recent galaxy observations show that star formation activity changes depending on galactic environments. In order to understand the diversity of galactic-scale star formation, it is crucial to understand the formation and evolution of giant molecular clouds in an extreme environment. We focus on observational evidence that bars in strongly barred galaxies lack massive stars even though quantities of molecular gas are sufficient to form stars. In this paper, we present a hydrodynamical simulation of a strongly barred galaxy, using a stellar potential which is taken from observational results of NGC1300, and we compare cloud properties between different galactic environments: bar, bar-end and spiral arms. We find that the mean of cloud's virial parameter is $\alpha_{\mathrm{vir}} \sim 1$ and that there is no environmental dependence, indicating that the gravitationally-bound state of a cloud is not behind the observational evidence of the lack of massive stars in strong bars. Instead, we focus on cloud-cloud collisions, which have been proposed as a triggering mechanism for massive star formation. We find that the collision speed in the bar is faster than those in the other regions. We examine the collision frequency using clouds' kinematics and conclude that the fast collisions in the bar could originate from random-like motion of clouds due to elliptical gas orbits shifted by the bar potential. These results suggest that the observed regions of lack of active star-formation in the strong bar originate from the fast cloud-cloud collisions, which are inefficient in forming massive stars, due to the galactic-scale violent gas motion.
\end{abstract}




\section{Introduction}


Galactic-scale star formation appears to have a wide variety depending on galactic-scale environments. Recent observational works have shown that systematic variations exist in a power-law relation between the gas surface density and the surface density of the star formation rate (SFR), and that the star formation activity is sensitive to global structural variations such as galaxy type \citep[e.g.][]{DaddiEtAl2010, LeroyEtAl2013}, conditions in the galactic central region \citep[e.g.][]{OkaEtAl2001} and the grand design spiral arms \citep[e.g.][]{ShethEtAl2002, MomoseEtAl2010}. In order to understand the diversity of galactic-scale star formation, understanding the formation and evolution of giant molecular clouds (GMCs), which are stellar nurseries, in extreme environments is quite informative.

A galactic bar structure has been known as an environment where the star formation is suppressed. The SFR in the bar is lower than that in the spiral arms, even when the gas surface density is almost the same \citep{MomoseEtAl2010, HirotaEtAl2014}; i.e., the star formation efficiency (SFE $ = \Sigma_{\mathrm{SFR}} / \Sigma_{\mathrm{gas}}$) is very low in the bar environment. It has been proposed that a strong shock and/or shear along the bar would inject turbulence into the dense gas to prevent star formation \citep{Tubbs1982, Athanassoula1992, DownesEtAl1996, ReynaudDownes1998, SoraiEtAl2012, MeidtEtAl2013}. However, the spatial scale of the gas stream in the bar is of the order of a kpc, which is much larger than that of individual GMCs ($< 50$ pc), let alone cloud cores ($< 0.1$ pc). Although we would expect that the large scale turbulence injected by the galactic bar could cascade down to smaller scales and affect gas structures to some extent, the actual physical process which prevents star formation is still not clear.

Recent high resolution hydrodynamical simulations have been a powerful tool for investigating the GMC's formation and their time evolution. The two-dimensional models of the weakly barred galaxy M83 have found that the SFE is lower than that of the spiral arms due to the strong internal turbulence of the clouds \citep{NimoriEtAl2013}. However, the three-dimensional simulations of the same galaxy have shown counter-evidence suggesting that the typical internal cloud velocity dispersion has little variation between clouds in the bar and spiral arm \citep{FujimotoEtAl2014}. Instead, \citet{FujimotoTaskerHabe2014} put a spotlight on a triggered star formation induced by cloud-cloud collisions. 

Interactions between GMCs can trigger a shock at the collision interface, which fragments into stars. The triggered star formation by cloud-cloud collisions has been suggested as a mechanism to create massive stars and super star clusters \citep[e.g.][]{HabeOhta1992, FurukawaEtAl2009, OhamaEtAl2010, FukuiEtAl2014, 2016ApJ...820...26F, 2018ApJ...859..166F, 2017ApJ...835..142T}. \citet{TakahiraTaskerHabe2014} performed hydrodynamical simulations of cloud-cloud collisions, and found that production of massive star-forming cores in a collision strongly depends on the relative speed of the two clouds. A fast collision shortens the gas accretion phase of the cloud cores formed, leading to suppression of core growth and massive star formation \citep[see also][]{TakahiraEtAl2018}. \citet{FujimotoTaskerHabe2014} found that clouds formed in the bar region typically collide faster than those in the spiral arms due to the elongated gas motion by the bar potential and that the high-velocity cloud-cloud collisions are responsible for the low SFE in the bar.

\begin{figure*}
	\includegraphics[width=160mm]{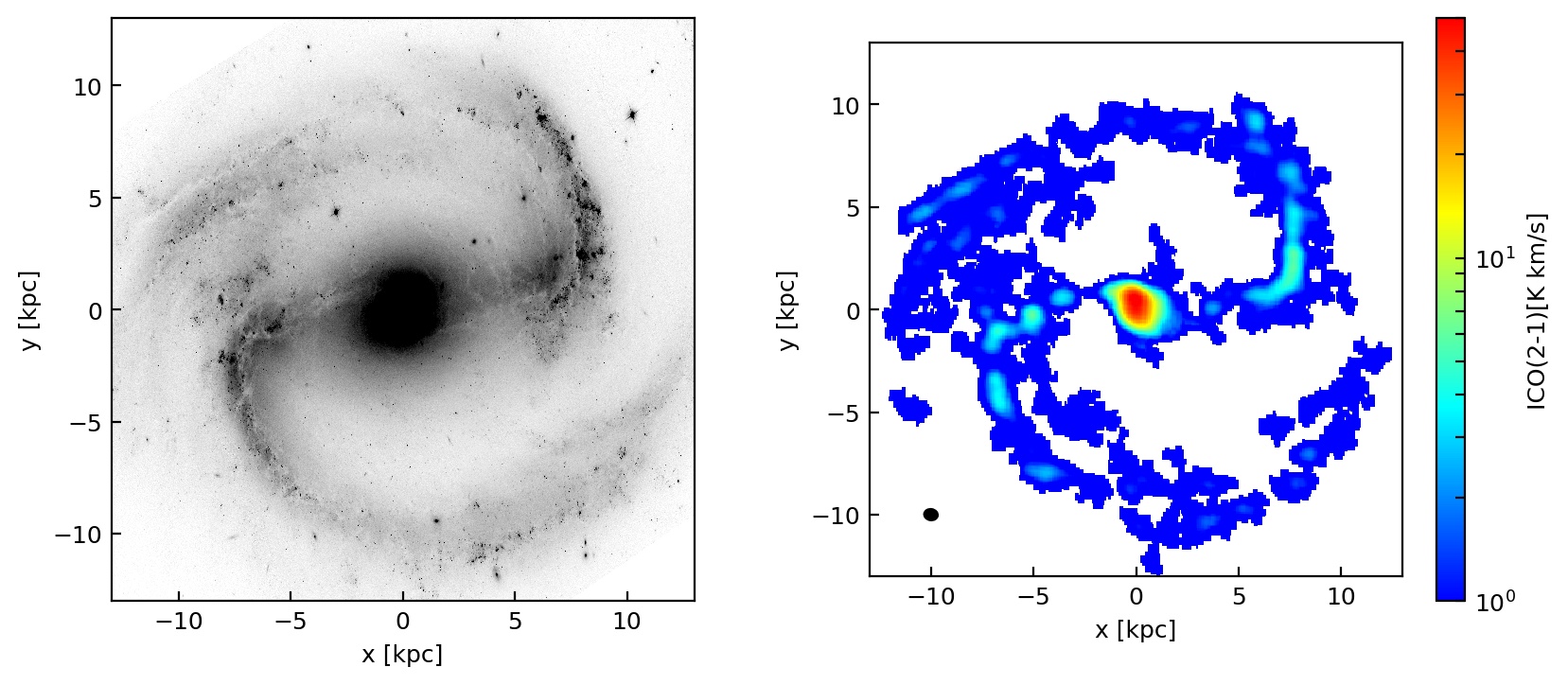}
    \caption{  
    Face-on views of V-band (F555W filter) image taken with HST (left) and velocity-integrated $^{12}$CO($2-1$) intensity map (right) corrected for the position angle and inclination of $-85.5^\circ$ and $50.2^{\circ}$ \citep{England1989a}, respectively. 
    The V-band image is obtained from the Hubble Legacy Archive (https://hla.stsci.edu/).  
    The $^{12}$CO($2-1$) image is generated from the ALMA archival data under project 2015.1.00925.S as proposed by B. Guillermo et al. We used the significant emission defined as follows: We first identify pixels with ${\rm S/N} > 4$ in at least two adjacent velocity channels. Next, we grow these regions to include adjacent pixels with ${\rm S/N} > 1.5$. The beamsize in the face-on view is $\rm 766~pc \times 645~pc$, which is represented as a black ellipse in bottom left corner.}
\label{fig:CO21_faceon}
\end{figure*}

The previous studies in terms of the star formation efficiency, however, have been made for barred galaxies with a rather weak bar. In order to unveil the cause for the low star formation efficiency in the bar region clearly, studying a barred galaxy with a strong bar is ideal, where the symptom of absence of star formation is most clearly seen. An extremely unproductive region for massive star formation is seen in such a strong bar; prominent H\textsc{ii} regions are mostly not seen although there are remarkable dust lanes along the stellar bar. Towards nearby strongly barred galaxies NGC1300 and NGC5383, sensitive CO observations were recently made and a large amount of molecular gas is proven to be present in the bar, even though they are not associated active star-forming regions \citep{MaedaEtAl2018}.

In this paper, we extend the previous studies of a weakly barred galaxy by \citet{FujimotoEtAl2014} and \citet{FujimotoTaskerHabe2014} to the strongly barred galaxy NGC1300 (the V-band image and $^{12}$CO($2-1$) intensity map are shown in Figure~\ref{fig:CO21_faceon}), and investigate the effects of the strong bar on the formation and evolution of GMCs to understand the physical mechanism of the observed lack of active star-forming regions in the strong bar. It is worth emphasizing here that unveiling the cause for the massive star formation suppression is also very important to understand the star formation in high redshift galaxies. For instance, high speed cloud collisions are expected to happen more frequently in the early universe, then the star formation may be suppressed, which would reduce the overproduction of stars without invoking the various feedback mechanisms proposed.

This paper is organised as follows. In Section~\ref{sec:Numerical Methods}, we present our model of the strongly barred galaxy NGC1300 and discuss details of the three-dimensional hydrodynamical simulation. Section~\ref{sec:Results} details our results, discussing first the global evolution of the galactic disc and moving on to exploring cloud properties and cloud-cloud collisions. In Section~\ref{sec:Discussion}, we make the discussion of the physical mechanism that causes the unproductive massive star formation in the strong bar more quantitatively by examining the collision frequencies of clouds using clouds' kinematics. Section~\ref{sec:Conclusions} presents our conclusions.

\section{Numerical Methods}
\label{sec:Numerical Methods}


\subsection{The code}
\label{sec:The code}

The simulations presented in this paper are of an isolated galaxy disc run using the adaptive mesh refinement (AMR) hydrodynamics code \textsc{enzo} \citep{BryanEtAl2014, BrummelEtAl2019}. The evolution of the gas is performed using a three-dimensional version of the \textsc{zeus} hydrodynamics algorithm \citep{StoneNorman1992}, which uses an artificial viscosity as a shock-capturing technique with the variable associated with this, the quadratic artificial viscosity, set to the default value of 2.0. 

The gas cools radiatively to 10 K using a one-dimensional cooling curve created from the \textsc{cloudy} package's cooling table for metals and \textsc{enzo}'s non-equilibrium cooling rates for atomic species of hydrogen and helium \citep{AbelEtAl1997, FerlandEtAl1998}. This is implemented as tabulated cooling rates as a function of density and temperature \citep{JinEtAl2017}. The Solar values for the metallicity and the dust-to-gas ratio are assumed. In addition to radiative cooling, the gas can also be heated via diffuse photoelectric heating in which electrons are ejected from dust grains via far-ultraviolet (FUV) photons. This is implemented as a constant heating rate of $8.5 \times 10^{-26}$ $\mathrm{erg\ s^{-1}}$ per hydrogen atom uniformly throughout the simulation box. This rate is chosen to match the expected heating rate assuming a UV background consistent with the Solar neighbourhood value \citep{Draine2011}. The cosmic ray heating is not considered. Self-gravity of the gas is also implemented.

The galaxy is modelled in a three-dimensional box of $(128\ \mathrm{kpc})^3$ with isolated gravitational boundary conditions and periodic fluid boundaries. The root grid is $128^3$ with an additional nine levels of refinement, producing a minimum cell size of $\Delta x = $ 1.953125 pc. We refine a cell by a factor of 2 whenever the mass included in the cell exceeds 1000 $\mathrm{M_{\odot}}$, or whenever the Jeans length, $\lambda_{\mathrm{J}} = c_{\mathrm{s}}\sqrt{\pi / (G \rho)}$, drops below four cell widths, satisfying the \citet{TrueloveEtAl1998} criterion. To prevent unresolved collapse at the finest resolution level, a pressure floor is implemented that injects energy to halt the collapse once the Jeans length becomes smaller than four cells. Gas in this regime follows a $\gamma = 2$ polytrope, $P \propto \rho^{\gamma}$. In addition to the static root grid, we impose five levels of statically refined regions enclosing the whole galactic disc of 20 kpc radius and 2 kpc height. This guarantees that the circular motion of the gas in the galactic disc is well resolved, with a maximum cell size of 31.25 pc.

In order to study the formation process of the cold dense clouds formed from the hot/warm diffuse gas, we use a relaxation strategy; in the first 560 Myr, we run the simulation without the cooling nor heating of the gas. The maximum refinement level is seven ($\Delta x = $ 7.8125 pc). During this first period, which corresponds to roughly two rotations of the pattern speed, the gaseous galactic structures such as the bar and spiral arms form in accordance with the galactic potential, and the galactic disc then settles into a quasi-equilibrium state. From $t = 560$ Myr, we include the cooling and heating of the gas and additional two levels of refinement ($\Delta x = $ 1.953125 pc), that allows the diffuse gas to collapse to the cold dense clouds. In order to focus on the early evolutionary stage of the clouds, we select the snapshot at $t = 600$ Myr when we analyse a single snapshot, and when we discuss time-averaged behaviour, the average is drawn from the time interval between $t = 590 - 610$ Myr. In fact, the median value of our cloud age at $t = 600$ Myr is $\sim 5$ Myr, which is shorter than the observational estimates of cloud lifetimes \citep[10-40 Myr; e.g.][]{KawamuraEtAl2009, ChevanceEtAl2019}. 

In order to study the effects of the galactic structures such as the bar on the evolution of the gas clouds alone, there is no star formation or stellar feedback in this simulation.

\subsection{The structure of the galactic disc}
\label{sec:The structure of the galactic disc}

Our galaxy is modelled on the barred spiral galaxy, NGC1300, with the stellar potential and gas distribution taken from observational results. Figure~\ref{fig:CO21_faceon} shows the face-on views of V-band image taken with Hubble Space Telescope (HST) and velocity-integrated $^{12}$CO($2-1$) intensity map corrected for the position angle and inclination of $-85.5^\circ$ and $50.2^{\circ}$ \citep{England1989a}, respectively. To enhance the $^{12}$CO($2-1$) emitting regions, we show the integrated intensity in the velocity range where the significant CO emission is seen.

\subsubsection{Stellar potential}
\label{sec:Stellar potential}

\begin{figure}
	\includegraphics[width=\columnwidth]{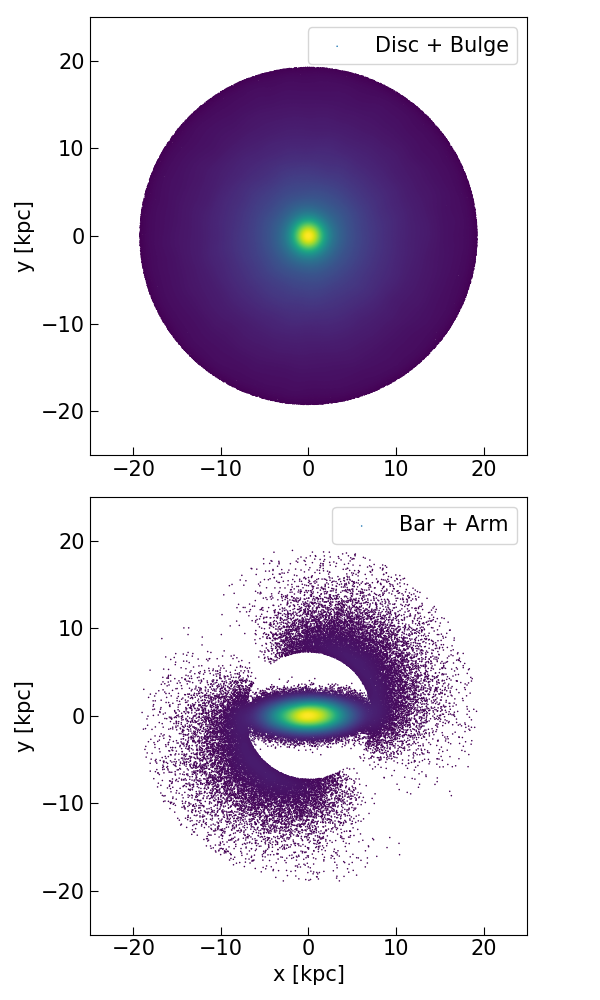}
    \caption{The stellar component of the galactic potential. Top panel shows the axisymmetric star particle distribution (disc and bulge). Bottom panel shows the non-axisymmetric star particle distribution (bar and arm), which rotates clockwise at a constant pattern speed.}
    \label{fig:particle_distribution}
\end{figure}

We use $10^6$ fixed-motion star particles to create a stellar potential model in keeping with the observed global characteristics of the stellar distribution in NGC1300. This model is based on the work of \citet{England1989b}, revised with the Hubble parameter of $H =73\ \mathrm{km\ s^{-1}\ Mpc^{-1}}$ and the distance to NGC1300 of $d = 20.7\ \mathrm{Mpc}$. The stellar density, consisting of the disc, bulge, bar and spiral arms, is given by
{\setlength\arraycolsep{2pt}
\begin{eqnarray}
    \rho_{\mathrm{star}}(r, \theta, z) &=& \Sigma (r, \theta) h(z) \nonumber \\
    &=& \{\Sigma_{\mathrm{disc}} (r) + \Sigma_{\mathrm{bulge}} (r) \nonumber \\
    && + \Sigma_{\mathrm{bar}} (r) \cos{(2 \theta)} + \Sigma_{\mathrm{arm}} (r) \sin{(2\theta)} \} h(z),
\end{eqnarray}
}
where $\Sigma_i (r)$ is the radial distribution of each component and $h(z)$ is the vertical distribution. Each of these are given by
\begin{equation}
    \Sigma_{\mathrm{disc}}(r)=\frac{M_{\mathrm{disc}}}{2\pi r_{\mathrm{disc}}^2}\exp \left(-\frac{r}{r_{\mathrm{disc}}}\right),
\end{equation}
\begin{equation}
    \Sigma_{\mathrm{bulge}}(r)=\frac{M_{\mathrm{bulge}}}{2\pi }\frac{r_{\mathrm{bulge}}}{\left(r^2 + r_{\mathrm{bulge}}^2\right)^{1.5}},
\end{equation}
\begin{equation}
    \Sigma_{\mathrm{bar}}(x, y)=\frac{M_{\mathrm{bar}}}{2\pi\sigma_x \sigma_y }\exp\left(-\frac{x^2}{2\sigma_x^2} - \frac{y^2}{2\sigma_y^2}\right),
\end{equation}
{\setlength\arraycolsep{2pt}
\begin{equation}
    \Sigma_{\mathrm{arm}}(r, \theta) = \left\{ \begin{array}{ll}
    \displaystyle \frac{M_{\mathrm{arm}}}{2\pi r_{\mathrm{arm}}^2 }\exp\left(-\frac{r}{r_{\mathrm{arm}}}\right)\left\{1+\sin (2\theta)\right\} &: r > 7.3\ \mathrm{kpc} \\
    & \\
    \displaystyle 0 &: \mathrm{otherwise},
    \end{array} \right.
\end{equation}
}
where $M_{\mathrm{disc}} = 6.78 \times 10^{10}\ \mathrm{M_{\odot}}$, $M_{\mathrm{bulge}} = 4.0 \times 10^{9}\ \mathrm{M_{\odot}}$, $M_{\mathrm{bar}} = 4.0 \times 10^{9}\ \mathrm{M_{\odot}}$, $M_{\mathrm{arm}} = 8.3 \times 10^{9}\ \mathrm{M_{\odot}}$, $r_{\mathrm{disc}} = 8.4$ kpc, $r_{\mathrm{bulge}} = 1.2$ kpc, $\sigma_x = 3$ kpc, $\sigma_y = 0.34 \sigma_x$ and $r_{\mathrm{arm}} = 2.4$ kpc. All these parameters are revised using the latest value of the Hubble parameter and the distance to NGC1300. We assume the vertical stellar distribution $h(z)$ is
\begin{equation}
    h(z) \propto \mathrm{sech}^2 \left( \frac{z}{200\ \mathrm{pc}} \right).
\end{equation}
Each star particle has a mass of $8.42 \times 10^4\ \mathrm{M_{\odot}}$, giving a total stellar mass of $M_* = 8.42 \times 10^{10}\ \mathrm{M_{\odot}}$.

Figure~\ref{fig:particle_distribution} shows the resulting distribution of the star particles. The top panel shows the axisymmetric star particles distribution, and the bottom panel shows the non-axisymmetric star particle distribution, which rotates at a constant pattern speed. We perform test simulations with four different pattern speeds: 16, 18, 20, and 22 $\mathrm{km\ s^{-1}\ kpc^{-1}}$. We select $\Omega =$ 20 $\mathrm{km\ s^{-1}\ kpc^{-1}}$ for the production run because the size of the gaseous bar formed in the hydrodynamic simulation matches the observational results for NGC1300. In both cases, the distance of the bar-end from the galactic centre is $7-8$ kpc (see Section~\ref{sec:Global structure and disc evolution}).

The average distance to the neighbouring particle is $\sim$ 50 pc at $r = 8$ kpc and $\sim$ 70 pc at $r = 16$ kpc. To remove the discreteness effects of the star particles, we smooth the particles' gravitational contribution by adding the mass on to the grid at AMR level 3, with a cell size of 125 pc. That is because the star particles should act as an external potential so that the gas feels gravity from the stellar component of the galaxy as a whole. Unlike an N-body galaxy simulation which includes gravitational interactions between particles, our fixed-motion star particles do not interact with each other, and they are put as a gravitational source only for the gas. To treat their gravitational effects as an external galactic potential, we smooth their gravitational contributions at a coarser level of the gas cell (the cell size is 125 pc), rather than putting them at the finest level whose cell size is 2 pc.

\subsubsection{Dark matter potential}
\label{sec:Dark matter potential}

In addition to the stellar potential, the galaxy sits in a static dark matter halo with an NFW model \citep{NavarroEtAl1996}. The halo concentration parameter is set to $c = 0.5$, and the virial mass of the halo (of which the mean density is 200 times the cosmological critical value) is set to $M_{200} = 5.4 \times 10^{10}\ \mathrm{M_{\odot}}$. These parameters are selected via comparisons between the rotation curve estimated from the stellar and dark matter potential and that from the observational results from NGC1300 \citep{England1989b}. We also compare and roughly match the rotation curve from the simulation with the observational results at around $r > 6$ kpc (see Section~\ref{sec:Global structure and disc evolution}).

\subsubsection{Gas mass}
\label{sec:Gas mass}

We estimate the H$_2$ gas mass from the archival $^{12}$CO($2-1)$ data which was observed with Atacama Compact Array (7m + total power) under project 2015.1.00925.S as proposed by B. Guillermo et al. We calibrated raw visibility data using the Common Astronomy Software Applications (CASA) and the observatory-provided calibration script. We imaged the interferometric map using the CLEAN algorithm in CASA by adopting Briggs weighting with robust $=$ 0.5. The resulting image is feathered with the total power image to recover the extended emission. 
The spatial resolution is $7.6^{\prime\prime} \times 4.1^{\prime\prime}$, corresponding to $760~{\rm pc} \times 410~{\rm pc}$ at the distance of $d =$ 20.7~Mpc. The rms noise of the data cube is $11.0~\rm mJy~beam^{-1}$ per 10.0 $\rm km~s^{-1}$ bin. The luminosity of the $^{12}$CO($2-1$) ($L_{\rm CO(2-1)}$) is $(3.0 \pm 0.3) \times 10^8~\rm K~km~s^{-1}~pc^2$. The total molecular gas mass is estimated to be $(1.70 \pm 0.78) \times 10^9~M_\odot$ by adopting the Galactic CO-to-H$_2$ conversion factor of 4.4 $M_\odot~(\rm K~km~s^{-1}~pc^2)^{-1}$ \citep{BolattoWolfireLeroy2013} and the line ratio ($I_{\rm CO(2-1)}/I_{\rm CO(1-0)}$) of 0.8 \citep{LeroyEtAl2009}. Combining the H$_2$ gas mass and the HI gas mass of $(3.72 \pm 0.60 ) \times 10^9\ \mathrm{M_\odot}$ taken from the observations by \citet{England1989a}, we estimate that the total gas mass in NGC1300 to be $(5.42 \pm 0.98) \times 10^9\ \mathrm{M_\odot}$.

\subsubsection{Initial gas distribution}
\label{sec:Initial gas distribution}

For the galaxy's radial gas distribution, we assume an initial exponential density profile with a radial scalelength of 8.4 kpc, based on the observations of \citet{England1989b}. The initial vertical distribution is assumed to be proportional to $\mathrm{sech}^2 (z/z_{\mathrm{h}})$ with a vertical scaleheight of $z_{\mathrm{h}} = 100$ pc.
The total gas mass of H\textsc{i} and H$_2$ in the simulation is $5.42 \times 10^9\ \mathrm{M_\odot}$ as described in Section~\ref{sec:Gas mass}. This gives an initial gas distribution:
\begin{equation}
\rho_{\mathrm{gas}}(r, z) = \rho_0 \exp{\left(- \frac{r}{8.4\ \mathrm{kpc}} \right)} \mathrm{sech}^2 \left( \frac{z}{100\ \mathrm{pc}}\right),
\end{equation}
where $\rho_0 = 9.152 \times 10^{-2}\ \mathrm{M_{\odot}\ pc^{-3}}$. The gas is set in circular motion as calculated via $V_{\mathrm{cir}}(r) = (GM_{\mathrm{tot}}/r)^{1/2}$, where $M_{\mathrm{tot}}$ is the enclosed mass of stars, dark matter and gas within the radius $r$.

\section{Results}
\label{sec:Results}

\subsection{Global structure and disc evolution}
\label{sec:Global structure and disc evolution}

\begin{figure*}
	\includegraphics[width=\hsize]{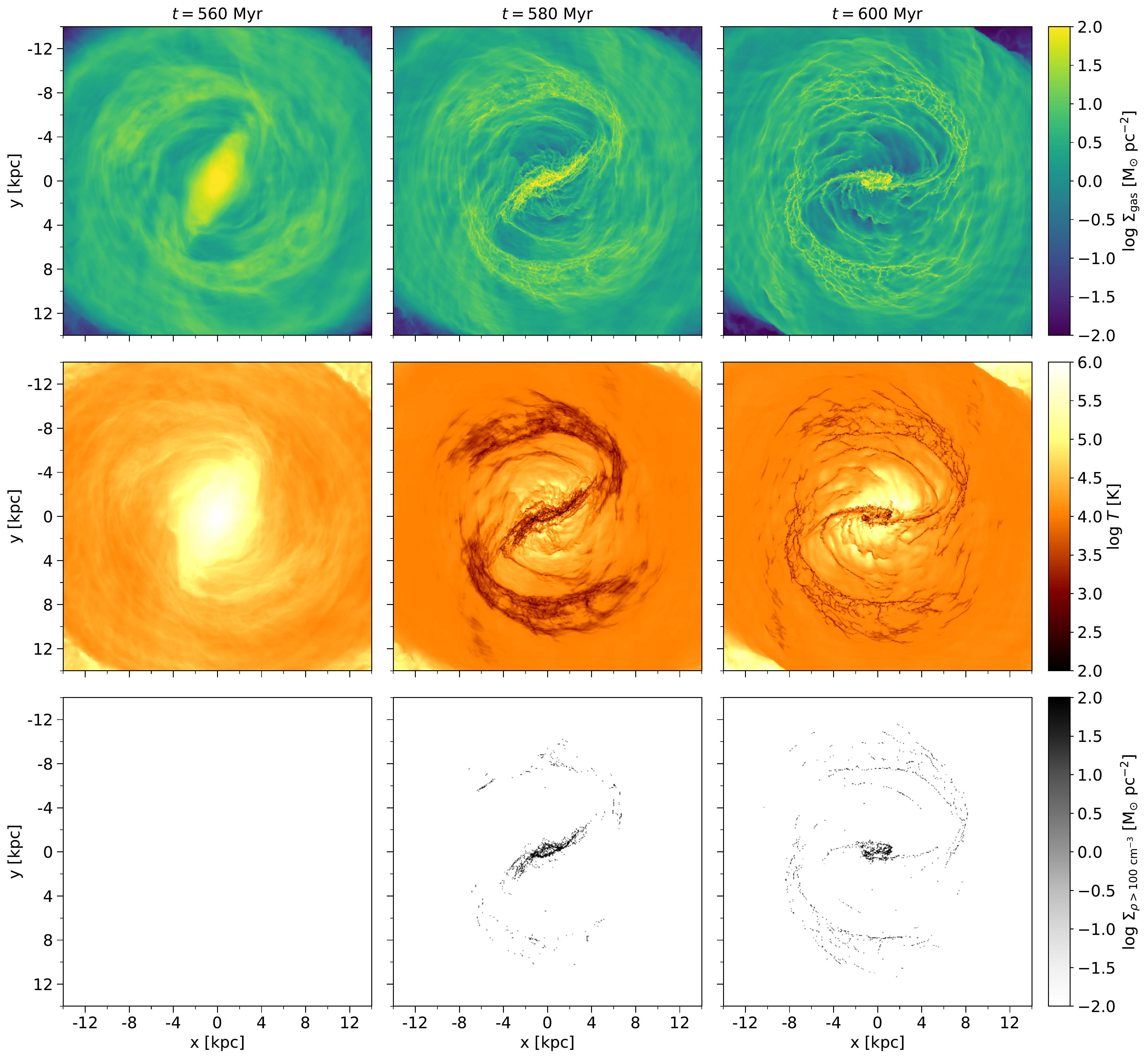}
    \caption{The global gas distribution in the face-on galactic disc. Left to right, the images show the disc at $t = $ 560, 580 and 600 Myr. Top to bottom, the images show the gas surface density, the density-weighted gas temperature and the surface density of dense gas of $\geq 100\ \mathrm{cm}^{-3}$. The galactic disc rotates clockwise.}
    \label{fig:galaxy_projections}
\end{figure*}

\begin{figure}
	\includegraphics[width=\columnwidth]{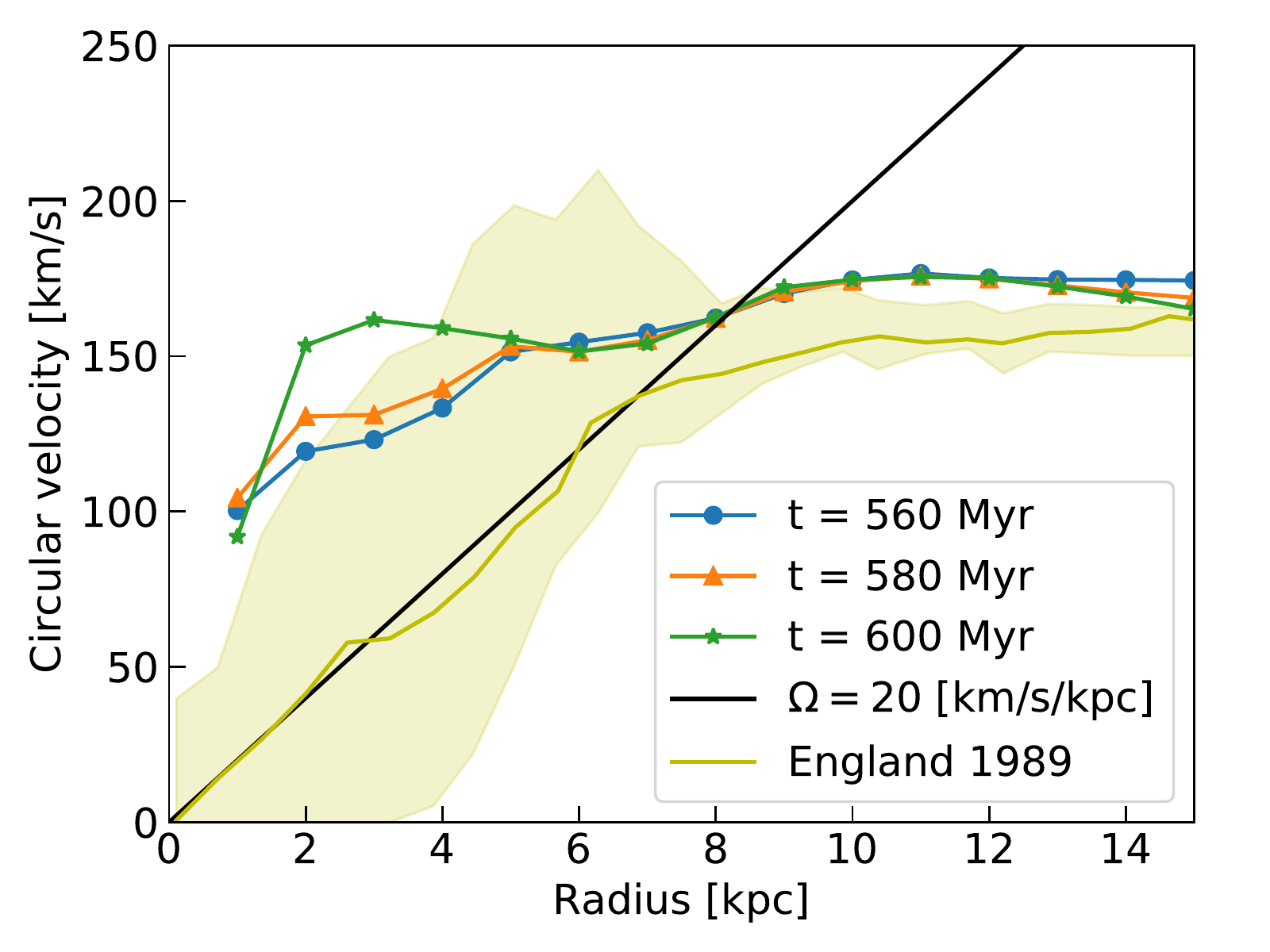}
    \caption{Azimuthally-averaged radial profiles of the gas circular velocity (mass-weighted average over $-1 < z < 1$ kpc) for the galactic disc at $t = $ 560, 580 and 600 Myr. The black solid line shows the constant pattern speed of the non-axisymmetric stellar potential. The dark yellow line shows the observed value from \citet{England1989b}, and their errors are shown with a filled area.}
    \label{fig:circular_velocity}
\end{figure}

In the initial stages of the simulation, the gas profile is smoothly exponential as described in Section~\ref{sec:Initial gas distribution}. As the simulation begins, the gas distribution changes according to the gravitational force by the stellar potential and the self-gravity, and the grand design pattern appears in the gas distribution. After $t = $ 150 Myr (a half pattern rotation period), the gaseous bar settles into a quasi-equilibrium with no large structural change. The grand design two spiral arms form and dissipate repeatedly by their outward expansion in a $200 \sim 300$ Myr period.

We include the cooling and heating of the gas from $t = $ 560 Myr, and then the gas fragments into knots and filaments. Figure~\ref{fig:galaxy_projections} shows the gas distribution in the galactic disc at three different times: $t = $ 560, 580 and 600 Myr. The top row is the gas surface density, the middle is the density-weighted temperature, and the bottom row is the surface density of the dense gas ($\geq 100\ \mathrm{cm}^{-3}$). At $t = $ 560 Myr, the galactic gas disc has already been stabilized, and the grand design of the gaseous bar and spiral arms has already formed. The temperature of the gaseous arms is around $10^4$ K, while the bar region has much higher temperature because of the shock heating by the strong elliptical gas motion in the stellar bar potential. The dense gas has not yet formed at that time. After the cooling of the gas is included, the gas cools down to several hundred K, and the gas begins to fragment into filaments and to form dense gas clouds as shown in the middle and right panels in Figure~\ref{fig:galaxy_projections}. 

The global galactic structures of the bar and spiral arms at $t = $ 600 Myr are similar to the $^{12}\mathrm{CO} (J = 2-1)$ image of NGC 1300 (see the right panel of Figure~\ref{fig:CO21_faceon}), with the bar-end at $r = 7 \sim 8$ kpc. The total dense gas mass ($\geq 100\ \mathrm{cm}^{-3}$) is $1.69 \times 10^9\ \mathrm{M_{\odot}}$, which is consistent with the observed molecular gas mass of $1.7 \times 10^9\ \mathrm{M_{\odot}}$ as shown in Section~\ref{sec:Gas mass}. The bar and spiral arms rotate in a clockwise sence in the panels with the non-axisymmetric stellar potential.

Figure~\ref{fig:circular_velocity} shows the radial profile of the mean circular velocity of the gas. This is calculated by using a mass-weighted average over $-1 < z < 1$ kpc. While time-dependent fluctuations are seen in the inner region of radius $r < 6$ kpc due to the strong elliptical gas motion inside the bar region, the profile shape remains unchanged in the large radius $r > 6$ kpc, keeping the circular velocity $150 \sim 160\ \mathrm{km\ s^{-1}}$, which is roughly consistent with the observed rotation velocities of NGC 1300
obtained from H\textsc{i} \citep{England1989b} and $^{12}\mathrm{CO} (J = 2-1)$.

\subsection{Cloud definition and classification}
\label{sec:Cloud definition and classification}

\begin{figure}
	\includegraphics[width=\columnwidth]{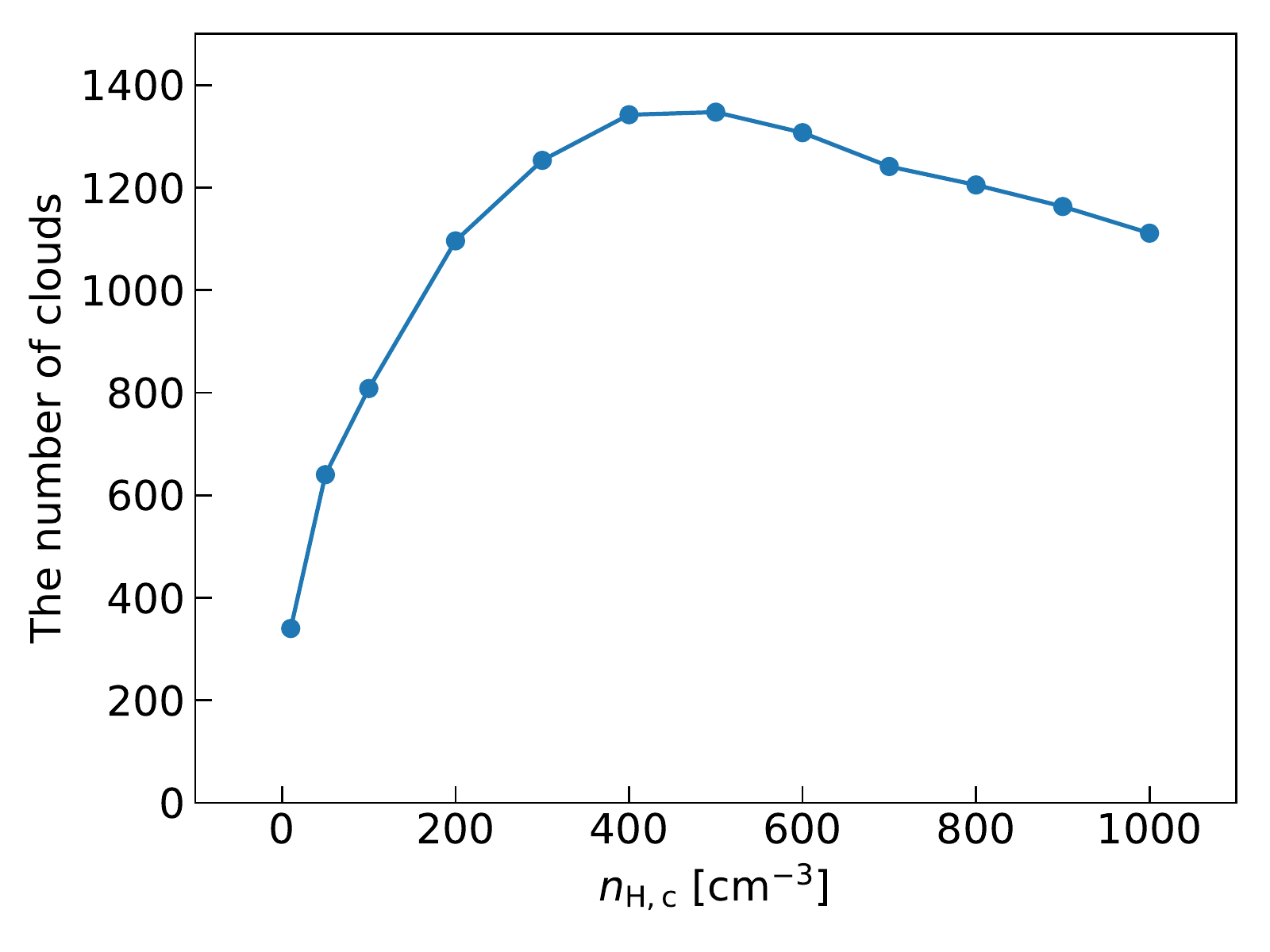}
    \caption{The total number of clouds identified in the simulated galactic disc, as a function of the threshold number density for cloud definition.}
    \label{fig:cloud_number}
\end{figure}

\begin{figure}
	\includegraphics[width=\columnwidth]{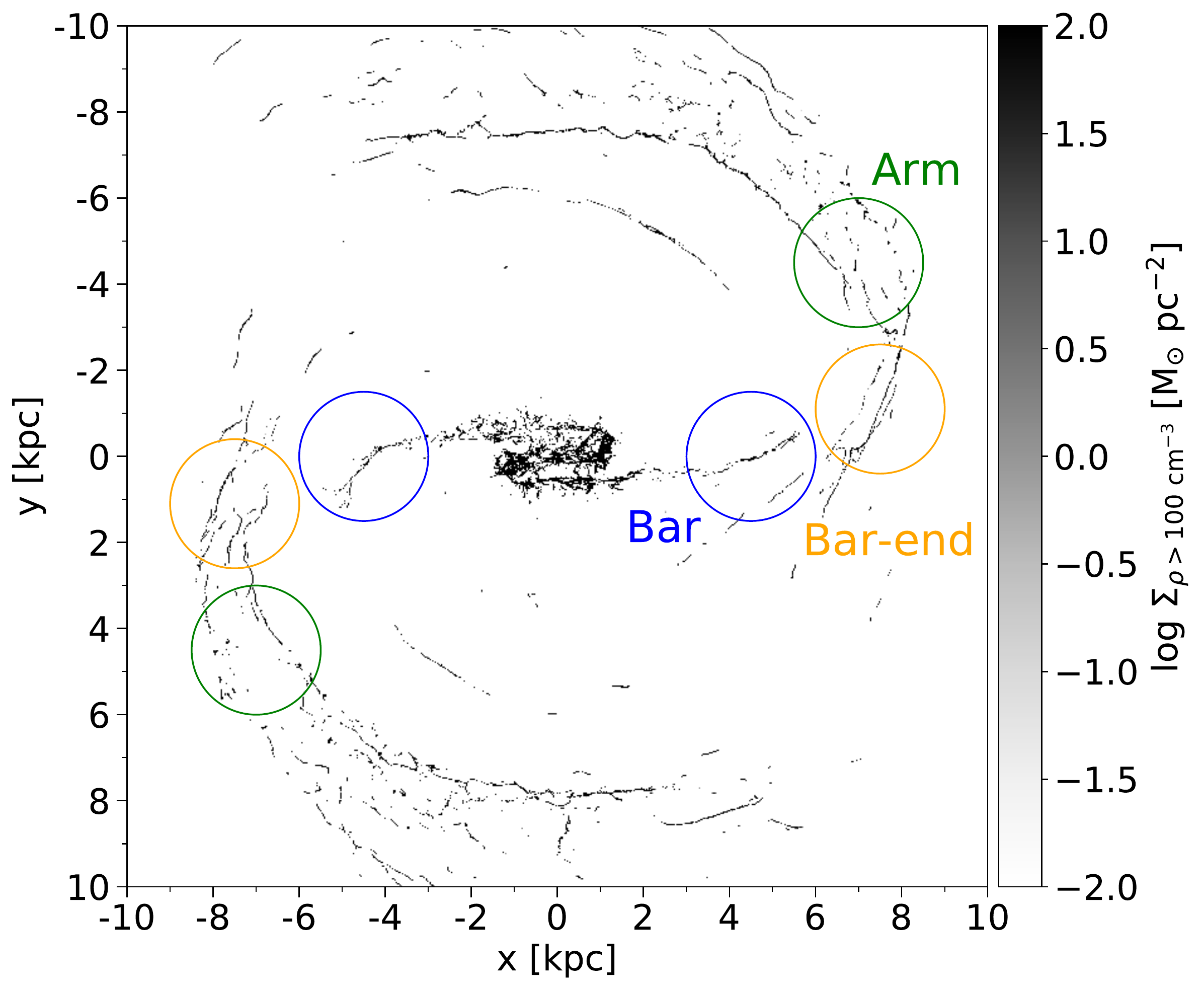}
    \caption{The three different galactic environments: \textit{Bar} (blue), \textit{Bar-end} (orange), \textit{Arm} (green). Each is a circle which is 3 kpc in diameter. The background colour shows the surface density of the dense gas of $\geq 100\ \mathrm{cm}^{-3}$ at $t = $ 600 Myr.}
    \label{fig:galactic_environments}
\end{figure}

We define a cloud in our simulation by a connected dense region that is enclosed by a contour at a threshold number density of H atom, $n_{\mathrm{H, c}}$; from the defined clouds in our simulation we exclude small clouds that contain fewer than $3^3$ cells, since we cannot properly resolve properties for such small clouds. Previous works have used $n_{\mathrm{H, c}} = 100\ \mathrm{cm}^{-3}$, because it is similar to the mean (volume-averaged) densities of typical Galactic GMCs \citep{TaskerTan2009, FujimotoEtAl2014, LiEtAl2018}. However, we use a slightly higher threshold density of $n_{\mathrm{H, c}} = 400\ \mathrm{cm}^{-3}$, since we can identify the maximum number of clouds at this threshold density. 
Figure~\ref{fig:cloud_number} shows the number of identified clouds with different threshold densities. When $n_{\mathrm{H, c}} < 400\ \mathrm{cm}^{-3}$, the number of identified clouds increases with increasing the threshold density, since there are clouds that are contained in larger structures at lower threshold densities. On the other hand, when $n_{\mathrm{H, c}} > 400\ \mathrm{cm}^{-3}$, the number of identified clouds decreases with the increase of the threshold density, since relatively diffuse low-density clouds are excluded from the cloud definition. Note that we do not see a significant difference in cloud properties, such as cloud mass and virial parameter (except for cloud radius), between $n_{\mathrm{H, c}} = $ 100 and 400 $\mathrm{cm}^{-3}$.

To compare the impact of different galactic environments on cloud properties, we define cloud groups based on the cloud's location in the disc by using three kinds of circles, the \textit{Bar}, \textit{Bar-end} and \textit{Arm}, of which positions at $t = $ 600 Myr are shown in Figure~\ref{fig:galactic_environments}. All circles co-rotate around the galactic centre with the same pattern speed of the bar and arm potential. In the snapshot at $t = $ 600 Myr, 99 clouds are in the \textit{Bar} region, 131 are in the \textit{Bar-end} region and 120 are in the \textit{Arm} region.

The three regions are completely different in terms of not only the location in the galactic disc but also the global-scale gas flow. As shown in Figure~\ref{fig:circular_velocity}, the gas in the \textit{Bar} region, where the galactic radius is $\ll$ 8 kpc, rotates faster than the non-axisymmetric bar potential. On the other hand, the gas in the \textit{Arm} region, where the galactic radius is $\gg$ 8 kpc, rotates slower than the spiral arm potential. The gas in the \textit{Bar-end} region, where the galactic radius is $\sim$ 8 kpc, co-rotates with the non-axisymmetric pattern.

\subsection{Cloud properties}
\label{sec:Cloud properties}

\begin{figure*}
	\includegraphics[width=\columnwidth]{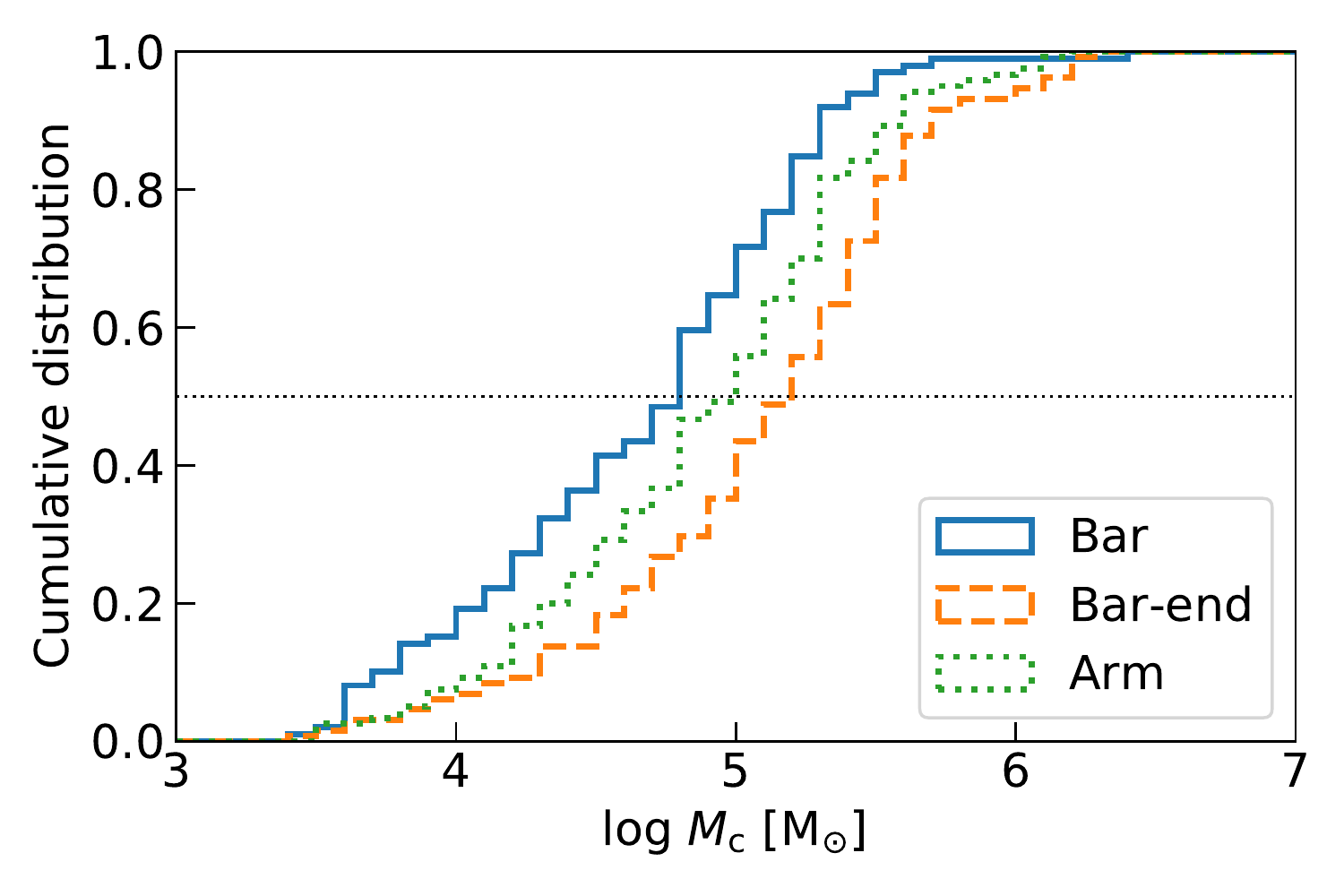}
	\includegraphics[width=\columnwidth]{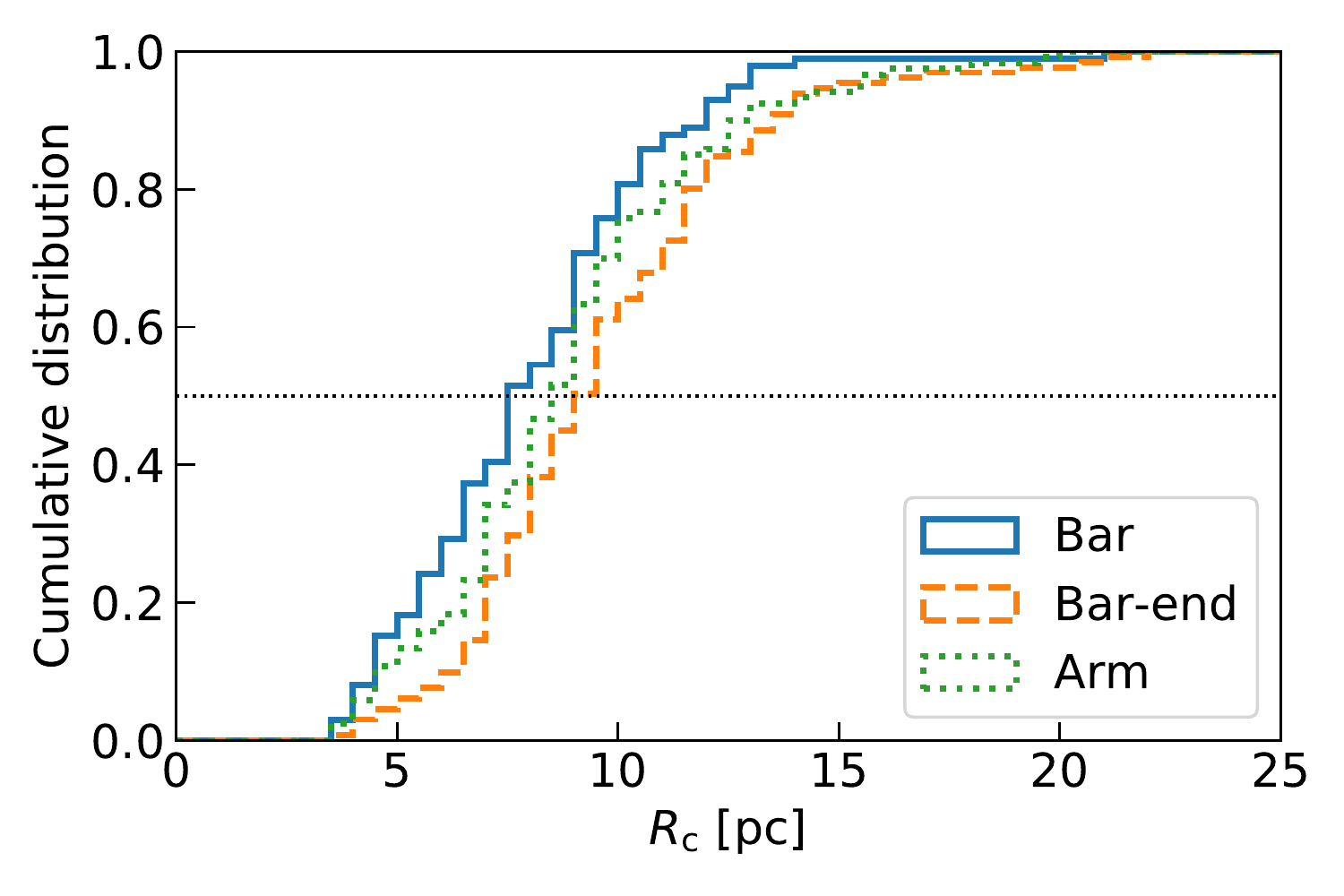}
	\includegraphics[width=\columnwidth]{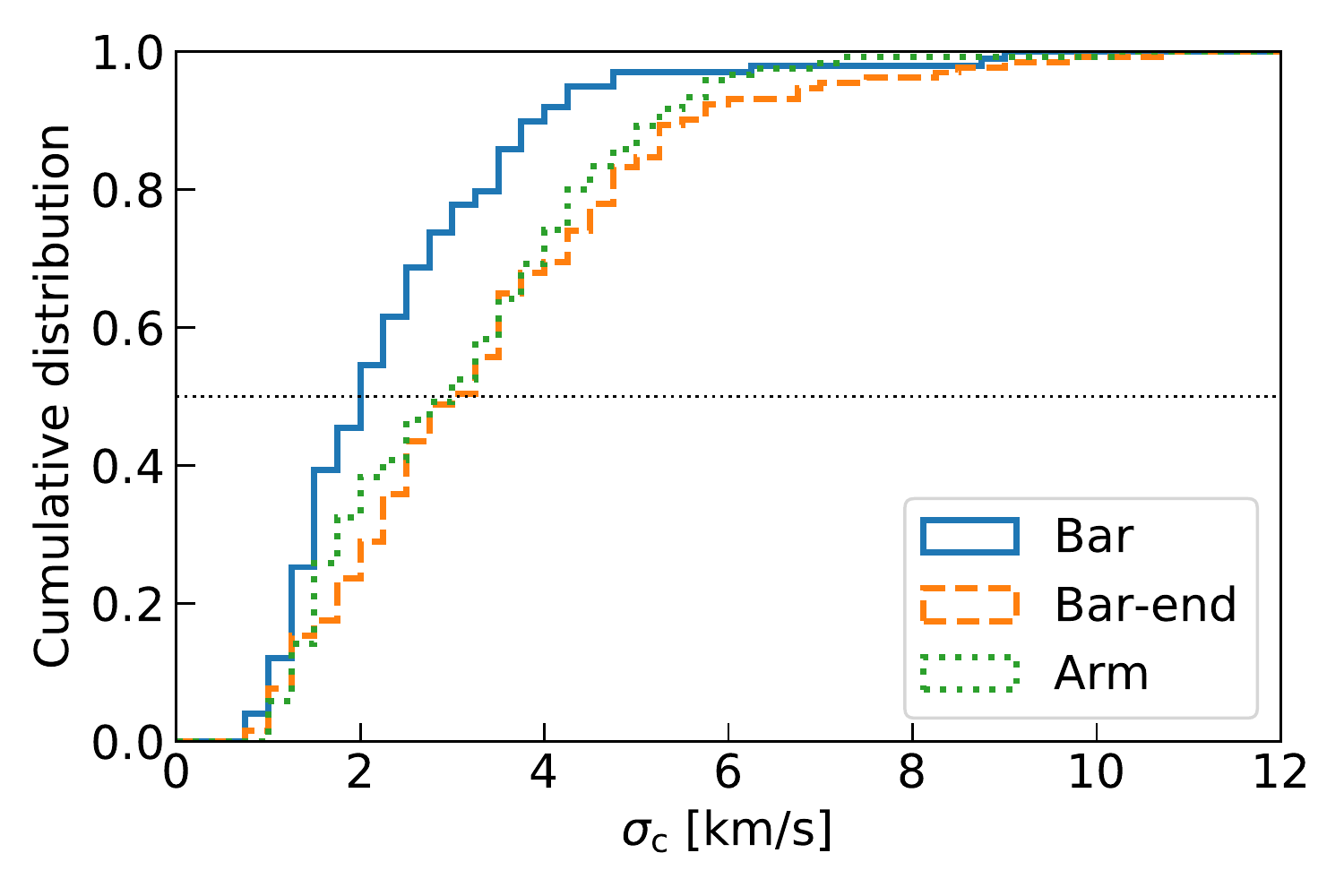}
	\includegraphics[width=\columnwidth]{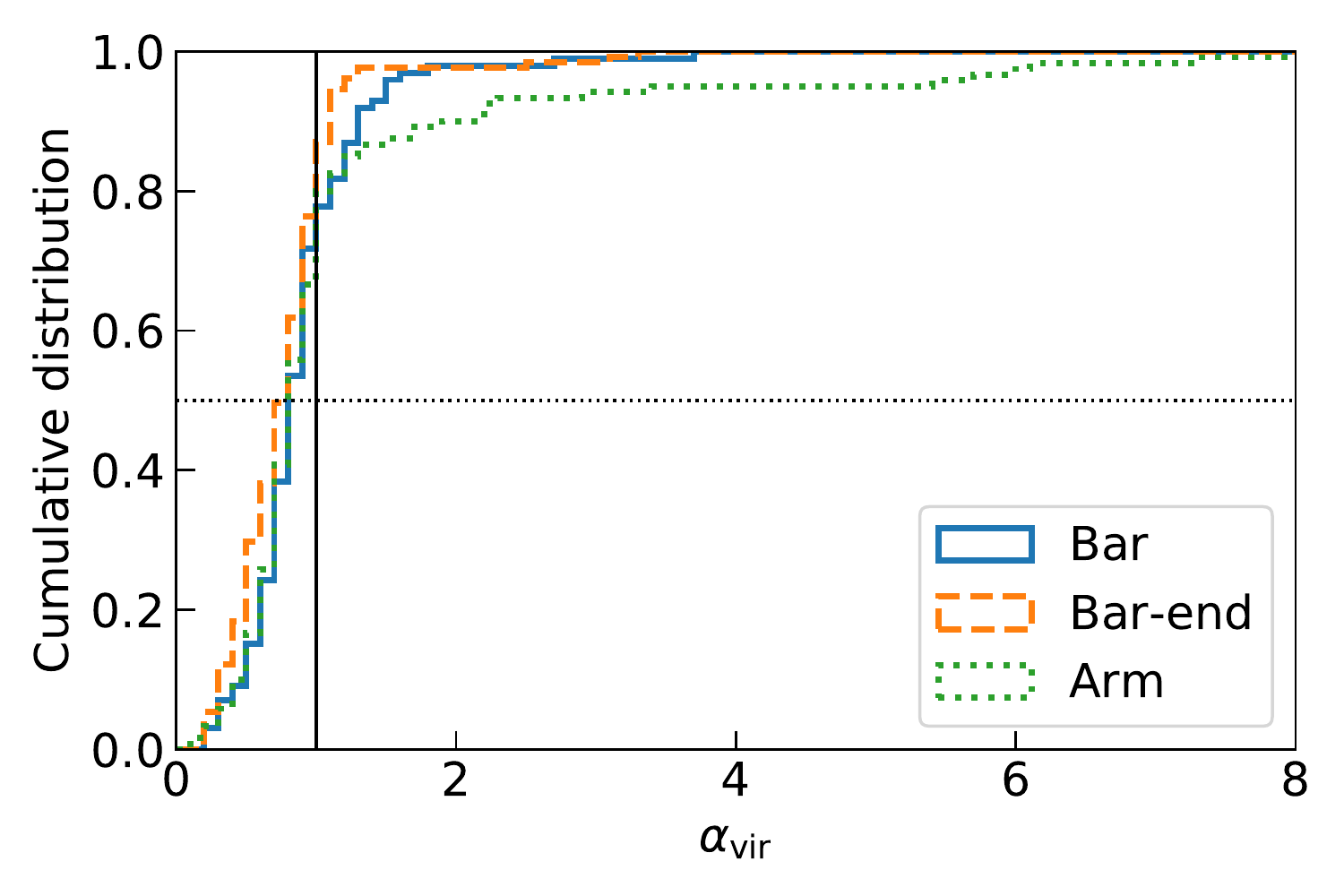}
    \caption{Normalized cumulative distribution function of cloud masses (top left), radii (top right), velocity dispersion (bottom left), and virial parameters (bottom right).}
    \label{fig:cloud_properties}
\end{figure*}

To see the impact of the galactic environment on the cloud formation, we plot the cloud property distributions for clouds in each of our defined regions at $t = $ 600 Myr in Figure~\ref{fig:cloud_properties}. The top left panel of Figure~\ref{fig:cloud_properties} shows the normalized cumulative distribution function of the cloud mass for these three environments, where the cloud mass is calculated by the sum of the total gas mass in each cell belonging to the cloud. In all three cases, the cloud mass ranges from $5 \times 10^3\ \mathrm{M_\odot} < M_{\mathrm{c}} < 5 \times 10^6\ \mathrm{M_\odot}$, and the median values lie at around $M_{\mathrm{c}} \simeq 10^5\ \mathrm{M_\odot}$, in reasonable agreement with the GMCs observed in nearby galaxies \citep[e.g.][]{RosolowskyEtAl2003, FreemanEtAl2017}. There is a slight environmental difference between the three regions; the \textit{Bar-end} clouds have the largest fraction of massive clouds, and, on the other hand, the \textit{Bar} clouds have the smallest. However, the difference is no more than 0.3 dex in log-scale.

The top right panel of Figure~\ref{fig:cloud_properties} shows the normalized cumulative distribution function of the cloud radius. We define the cloud radius as
\begin{equation}
    R_{\mathrm{c}} = \left(\frac{3 V_{\mathrm{c}}}{4 \pi} \right)^{1/3},
\end{equation}
where $V_{\mathrm{c}}$ is the cloud volume which is calculated the sum of the cell volumes in the cloud. In all three regions, the median values lie at around $R_{\mathrm{c}} \sim 8$ pc, and their shapes of the distribution are similar. 

The bottom left panel of Figure~\ref{fig:cloud_properties} shows the normalized cumulative distribution function of the velocity dispersion of the clouds. We define the velocity dispersion as
\begin{equation}
    \sigma_{\mathrm{c}} = \sqrt{\sigma_{\mathrm{1D}}^2 + c_{\mathrm{s}}^2},
\end{equation}
where $\sigma_{\mathrm{1D}}$ is the mass-averaged one-dimensional velocity dispersion defined as $\sigma_{\mathrm{1D}} = \langle\sqrt{|\vec{v} - \vec{v}_{\mathrm{CoM}}|^2/ 3}\rangle$, where $\vec{v}$ is the velocity of the gas and $\vec{v}_{\mathrm{CoM}}$ is the cloud's centre of mass velocity, $c_{\mathrm{s}}$ is the sound speed, and the angle brackets indicate a mass-weighted average over the cells in the cloud. Clouds in the \textit{Bar-end} and \textit{Arm} regions show median values of $\sim 3\ \mathrm{km\ s^{-1}}$, and the \textit{Bar} clouds have slightly smaller velocity dispersion of $\sim 2\ \mathrm{km\ s^{-1}}$. This might be because clouds are less massive in the \textit{Bar}. Since the cloud mass is small, their gravitational potential well can be shallow, which could result in failing to catch their surrounding high velocity dispersion gas.

The bottom right panel of Figure~\ref{fig:cloud_properties} shows the normalized cumulative distribution function of the cloud virial parameter defined as \citep{BertoldiMcKee1992},
\begin{equation}
    \alpha_{\mathrm{vir}} = \frac{5 \sigma_{\mathrm{c}}^2 R_{\mathrm{c}}}{G M_{\mathrm{c}}}.
\end{equation}
The distributions show that the median values are around $\alpha_{\mathrm{vir}} \sim 1$, indicating that most clouds are gravitationally bound, and that there is almost no environmental dependence. This result is consistent with \citet{MaedaEtAl2020} who identified GMCs in NGC 1300 by ALMA $^{12}$CO($1-0$) observations and found that there is no significant variation in the $\alpha_{\rm vir}$ between the arm and bar region. We have shown detailed comparison with the observation in Appendix~\ref{sec:Cloud properties: comparison with observations}.

We can see a high-end tail of the virial parameter in the \textit{Arm} region. When we make the distributions with smaller regions with 1.5 kpc diameter, the over-all distribution does not change, but the high-end tail disappears, indicating that it appears by chance in the plot. We also check histograms of their properties, and again, we do not find any clearer difference between three regions.

In summary, we do not find any significant dependence of cloud properties on the galactic regions, in particular, the cloud virial parameter. $\alpha_{\mathrm{vir}} \sim 1$ indicates that most clouds are gravitationally bound and that low- and intermediate-mass stars can form in all regions. In terms of massive stars, however, we still cannot explain the observed evidence of the lack of massive star formation in strong bars with the cloud's virial parameters because we find no environmental dependence. In fact, recent ALMA observations also show no significant variation in the virial parameter \citep{MaedaEtAl2020}. Instead, in the following sections, we focus on a cloud-cloud collision, which has been suggested as a triggering mechanism of massive star formation \citep[e.g.][]{HabeOhta1992, FurukawaEtAl2009, OhamaEtAl2010, FukuiEtAl2014, 2016ApJ...820...26F, 2018ApJ...859..166F, 2017ApJ...835..142T}, and look for the cause of the lack of massive stars in strong bars.

\subsection{Cloud-cloud collisions}
\label{sec:Cloud-cloud collisions}

\begin{table}
	\centering
	\caption{The number of cloud-cloud collisions, the number of clouds, the collision frequency per Gyr of clouds in each regions, and the surface density of the estimated SFR assuming that star formation can be triggered only by cloud-cloud collisions, and the observed SFR in NGC1300 (Maeda et al. in preparation).}
	\label{tab:collision_frequencies}
	\begin{tabular}{lccc} 
		\hline
		 & \textit{Bar} & \textit{Bar-end} & \textit{Arm}\\
		\hline
		The number of collisions: $n_{\mathrm{ccc}}$ & 28 & 46 & 33\\
		\hspace{20pt} ($t = 590 - 610$ Myr) \\
		The number of clouds: $n_{\mathrm{cloud}}$ & 99 & 131 & 120\\
		\hspace{20pt} ($t = 600$ Myr) \\
		Collision frequency: $f_{\mathrm{ccc}}$ & 14.1 & 17.6 & 13.8\\
		\hspace{20pt} $[\mathrm{Gyr}^{-1}]$ \\
		Estimated SFR: $\Sigma_\mathrm{SFR}$ & 3.38 & 5.52 & 2.87\\
		\hspace{20pt} $[10^{-3}\ \mathrm{M_{\odot}\ yr^{-1}\ kpc^{-2}}]$ \\
		Observed SFR: $\Sigma_\mathrm{SFR, Obs.}$ & $\ll$ 1.0 & 3.9 - 4.6 & 4.7 - 5.9\\
		\hspace{20pt} $[10^{-3}\ \mathrm{M_{\odot}\ yr^{-1}\ kpc^{-2}}]$ \\
		\hline
	\end{tabular}
\end{table}

To see the impact of galactic environments on the clouds' interactions, we follow the evolution of the clouds, analysing simulation outputs at intervals of 0.2 Myr between $t = $ 590 and 610 Myr and mapping the clouds between the outputs with a tag number assigned to each cloud. The algorithm of this cloud tracking is described in \citet{TaskerTan2009}. We define a collision if we find a single cloud near the positions predicted by two other clouds before 0.2 Myr. Table~\ref{tab:collision_frequencies} shows the number of cloud-cloud collisions occurred in between $t = 590 - 610$ Myr, the number of clouds at $t = 600$ Myr, the collision frequency per clouds per Gyr in each region, the surface density of the estimated star formation rates (SFRs), and the observed ones in NGC1300.

The collision frequency is defined as,
\begin{equation}
    f_{\mathrm{ccc}} = \frac{n_{\mathrm{ccc}}}{n_{\mathrm{cloud}} \Delta t},
\end{equation}
where $n_{\mathrm{ccc}}$ is the number of collisions, $n_{\mathrm{cloud}}$ is the number of clouds, and $\Delta t$ is the tracking time of 20 Myr. The collision frequencies are 10-20 $\mathrm{Gyr}^{-1}$, which is slightly lower than those in a Milky-Way-type galaxy. Collision frequencies of 30-40 $\mathrm{Gyr}^{-1}$ are estimated for Milky-Way-type galaxy simulations \citep{TaskerTan2009, DobbsPringleDuarte-Cabral2015}. Because the modelling galaxy is different, it is not surprising that we get different collision frequencies. The other possible reason is that our clouds are more compact and smaller because of the higher threshold density for cloud definition, which can make the cross-sections of clouds smaller and then the collision frequencies lower.

We see that there is no significant environmental difference in the collision frequency, suggesting that a frequency of the cloud-cloud collisions does not seem to be the cause of the lack of the massive stars in strong bars in observations.

For a more quantitative discussion, we estimate SFRs assuming that star formation can be initiated only by cloud-cloud collisions and compare the estimates with those of observations. Based on a cloud collision triggered star formation model proposed by \citet{Tan2000}, we define the surface density of the SFR as,
\begin{equation}
    \Sigma_{\mathrm{SFR}} = \epsilon f_{\mathrm{sf}} \left(\sum_{\Delta t} \sum_{A} M_{\mathrm{ccc}} \right) \bigg/ A \Delta t,
\end{equation}
where $\epsilon$ is the total star formation efficiency, $f_{\mathrm{sf}}$ is the fraction of cloud collisions which lead to star formation, $M_{\mathrm{ccc}}$ is the colliding cloud mass just after the incident, and $A$ is the surface area of each galactic region. For simplicity, we use the default values of $\epsilon = 0.2$ and $f_{\mathrm{sf}} = 0.5$ in \citet{Tan2000}. 

Again, we do not see a significant environmental dependence as with the collision frequencies. Compared to the observed SFRs in NGC1300 (Maeda et al. in preparation), although the estimates for the \textit{Bar-end} and \textit{Arm} are roughly consistent, the one for the \textit{Bar} is much larger than the upper limit of the observed value; the estimated SFR in the \textit{Bar} is $3.38 \times 10^{-3}\ \mathrm{M_{\odot}\ yr^{-1}\ kpc^{-2}}$, and the observed one is less than $10^{-3}\ \mathrm{M_{\odot}\ yr^{-1}\ kpc^{-2}}$. \citet{FujimotoTaskerHabe2014} suggested that using the fixed values for $\epsilon$ and $f_{\mathrm{sf}}$ is too simplistic because not all cloud-cloud collisions should have the same efficiency to form stars, and the efficiency should be changed with properties of the collisions, such as the collision velocity. Hydrodynamical simulations of a cloud-cloud collision have shown that there is a clear dependence of the production rate of massive star-forming cores on the relative speed of the two colliding clouds \citep{TakahiraTaskerHabe2014}.

\begin{figure}
	\includegraphics[width=\columnwidth]{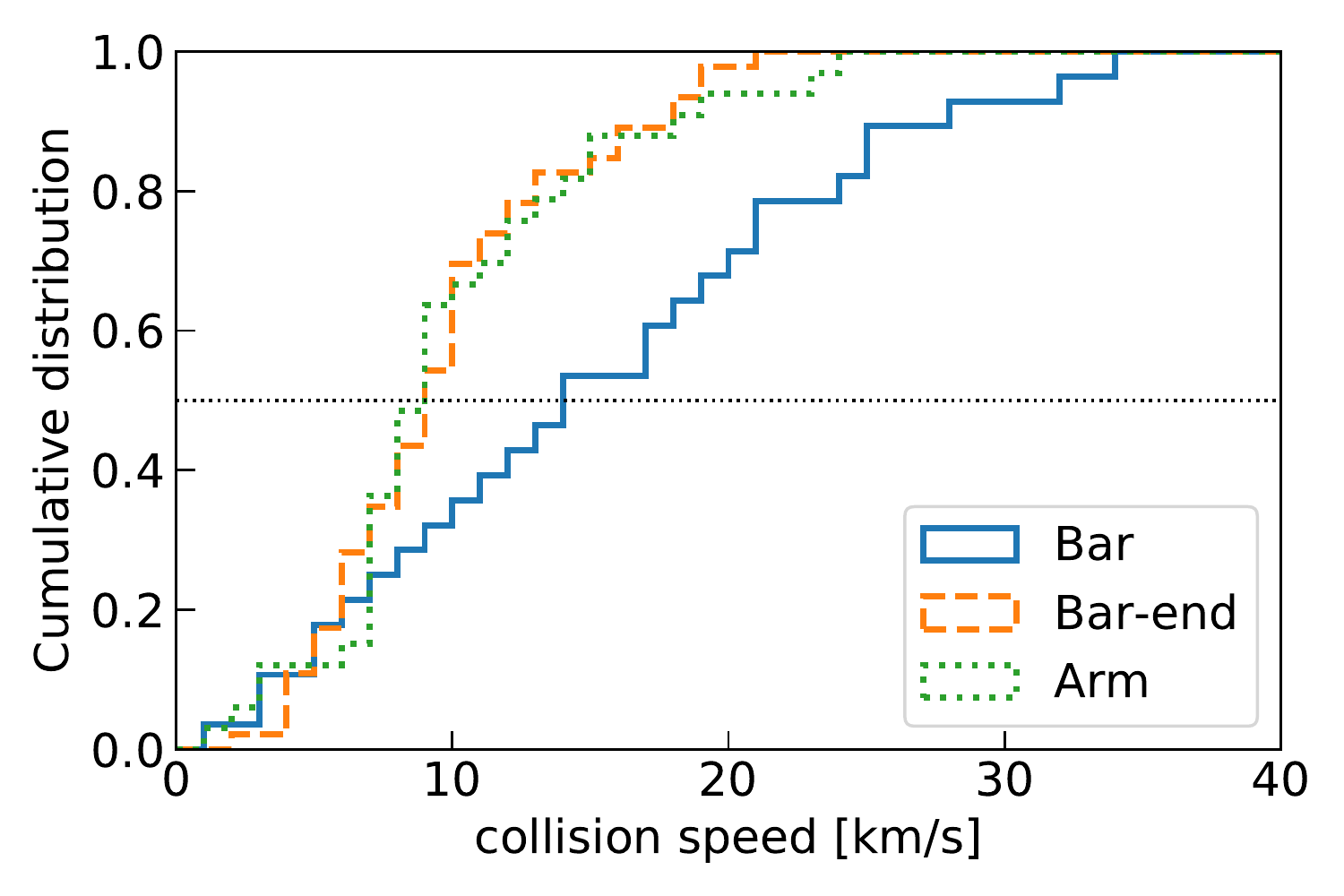}
    \caption{Normalized cumulative distribution function of the cloud-cloud collision speed. The samples are the collisions occurred in 20 Myr between $t = $ 590 and 610 Myr, as shown in Table~\ref{tab:collision_frequencies}.}
    \label{fig:collision_speed}
\end{figure}

We find that there is a clear environmental dependence in the collision speed. Figure~\ref{fig:collision_speed} shows the normalized cumulative distribution function of the cloud-cloud collision speed. The samples are the collisions occurred in 20 Myr between $t = $ 590 and 610 Myr, as shown in Table~\ref{tab:collision_frequencies}. Collisions in the \textit{Bar} have the median of $\sim 15\ \mathrm{km\ s^{-1}}$ while collisions in the \textit{Bar-end} and \textit{Arm} have $\sim 10\ \mathrm{km\ s^{-1}}$, which is $5\ \mathrm{km\ s^{-1}}$ smaller than that in the \textit{Bar}. Fraction of collided clouds with collision speed more than 20 km/s is roughly 40 per cent in the \textit{Bar} region, and, on the other hand, less than 10 per cent in the other regions. Such a high-speed cloud-cloud collision can shorten the gas accretion phase of the cloud cores formed, leading to suppression of core growth and massive star formation \citep[][]{TakahiraTaskerHabe2014, TakahiraEtAl2018}, suggesting that the fast cloud-collisions is the cause of the lack of massive stars in strong bars.

\begin{figure}
    \includegraphics[width=\columnwidth]{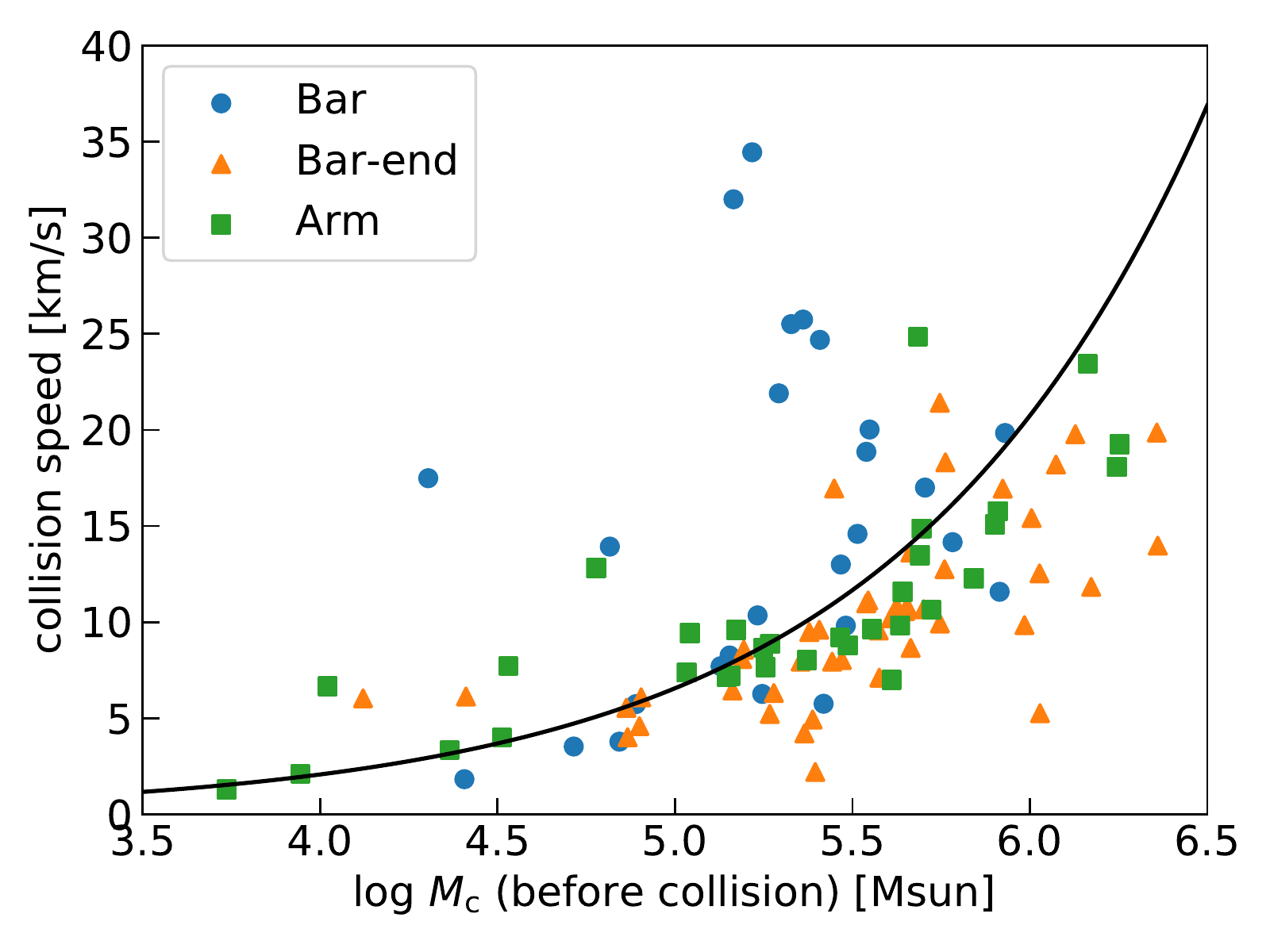}
    \caption{The scatter plots of the cloud-cloud collision speed versus the mass of the most massive cloud in each collision. The samples are the collisions occurred in 20 Myr between $t = $ 590 and 610 Myr, as shown in Table~\ref{tab:collision_frequencies}. The black line shows a free-fall speed $v_{\mathrm{ff}}$ as a function of cloud mass.}
    \label{fig:collision_scatter_plots}
\end{figure}

The fast collisions in the \textit{Bar} do not correlate with cloud mass. Figure~\ref{fig:collision_scatter_plots} shows the scatter plot between the collision speed and the mass of the most massive cloud in each collision. There is a rough correlation between them in the \textit{Bar-end} and \textit{Arm}, suggesting that the cloud collisions in these regions could be caused by gravitational interactions between clouds. On the other hand, the \textit{Bar} clouds have large scatter. We also investigate correlation between the collision speed and the cloud mass ratio between the colliding partners, and find no correlation between them.

For a more quantitative discussion, we overplot the free-fall speed as a function of cloud mass with a black line. The free-fall speed is defined as,
\begin{equation}
    v_{\mathrm{ff}} = \sqrt{2 \mathrm{G} M_{\mathrm{c}} \left( \frac{1}{2 R_{\mathrm{c}}} - \frac{1}{D_{\mathrm{nearest}}} \right) }, 
\end{equation}
where $D_{\mathrm{nearest}}$ is the distance to the nearest cloud. For simplicity, we use the median values: $R_{\mathrm{c}} = $ 8 pc and $D_{\mathrm{nearest}} = $ 100 pc. 

It shows that although there is a small deviation, the \textit{Bar-end} and \textit{Arm} clouds roughly fit the free-fall line, supporting the idea of a large contribution of gravitational interactions between clouds to driving collisions. We note that not only gravitational interactions but also other physical mechanism, such as a hydrodynamical drag from the low-density surrounding gas and a coagulation of the ISM around the spiral arm and the resonance point produced by the global galactic potential, could also be the cause of the cloud collisions. That could be the reason we see a small deviation in the scatter plot in those regions.

On the contrary to the \textit{Bar-end} and \textit{Arm}, a large fraction of clouds in the Bar region have much faster collision speed than the free-fall speed, indicating that the gravitational interactions have only a small contribution to driving collisions.

\begin{figure*}
	\includegraphics[width=\hsize]{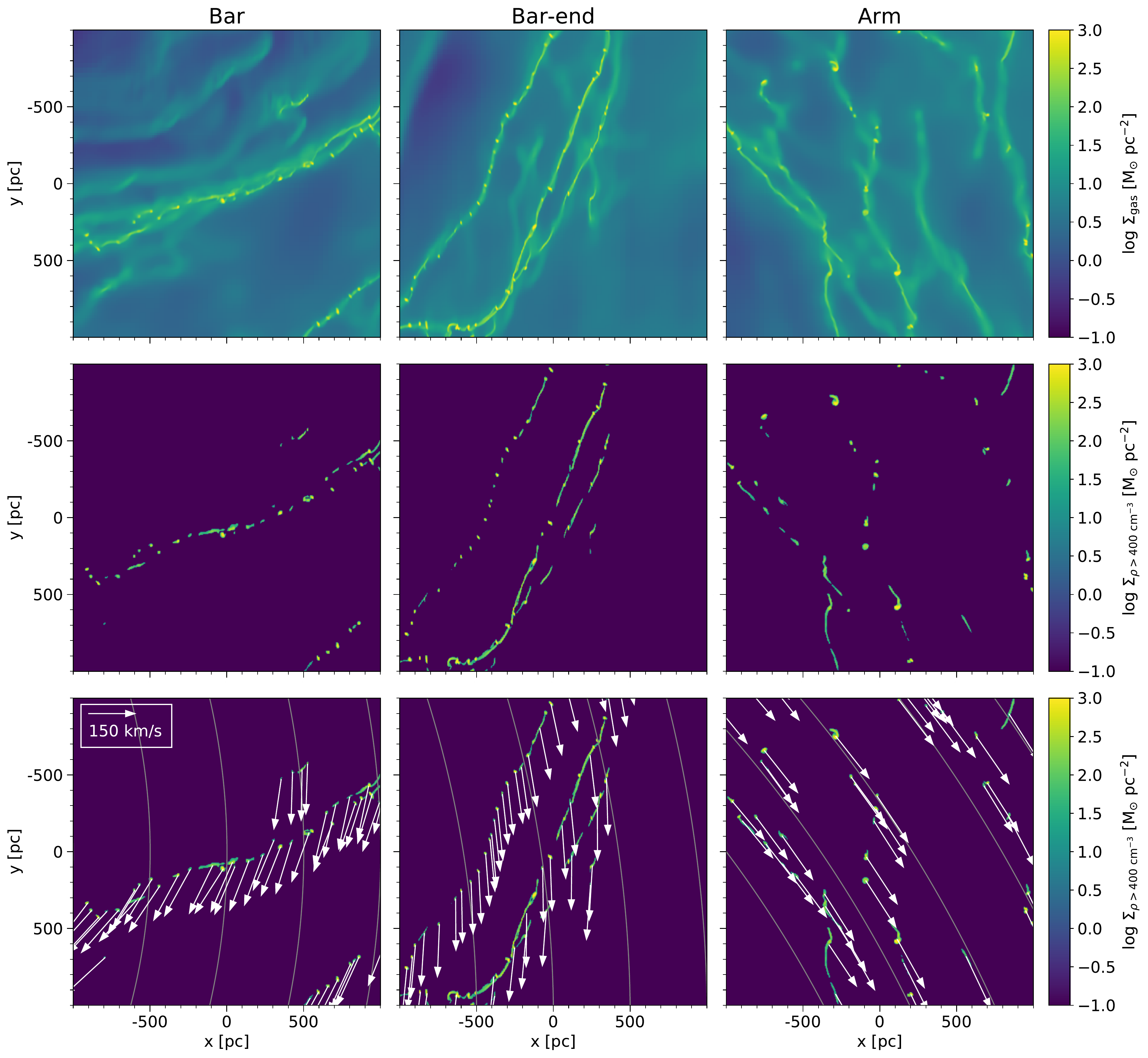}
    \caption{Zoomed-in on each galactic environments on the right side of the disc at $t = $ 600 Myr. Left to right, the images show the \textit{Bar}, \textit{Bar-end} and \textit{Arm} regions. Top to bottom, the images show the gas surface density, the dense gas surface density of $\geq 400\ \mathrm{cm}^{-3}$ and the same plot but overlaid with arrows showing the cloud's bulk projected velocity. The arcs in the bottom images show galactocentric radii at intervals of 500 pc.}
    \label{fig:zoom_projections}
\end{figure*}

We can confirm the origin of the high-speed collisions in the \textit{Bar} region using visual inspection of the galactic disc. Figure~\ref{fig:zoom_projections} shows 2 kpc patches of the total (top row) and dense (middle and bottom row) gas surface density in the \textit{Bar}, \textit{Bar-end}, and \textit{Arm} regions. The bottom panels show the projected bulk velocities of the clouds by arrows.

The bottom panels of Figure~\ref{fig:zoom_projections} show that motion of clouds in the \textit{Bar} region deviates from the galactic circular motion. This is due to the elongated elliptical gas motion shifted by the galactic bar potential. On the other hand, clouds in the \textit{Arm} regions are close to the circular orbit around the galactic centre. Clouds in the \textit{Bar-end} are moderate. The elongated motion of the gas in the \textit{Bar} could cause the high speed collisions of clouds, as we have shown that some of the velocity are crossing each other, as seen around at $(x, y) = (-700, 400)$ pc. Such irregular cloud velocities are hardly seen in the \textit{Bar-end} and \textit{Arm} regions. In Appendix~\ref{sec:Example of fast collision}, we have shown one example of the fast collision induced by a filament-filament collision due to the elongated gas motion in the \textit{Bar} region.

\begin{figure}
    \centering
    \includegraphics[width=\columnwidth]{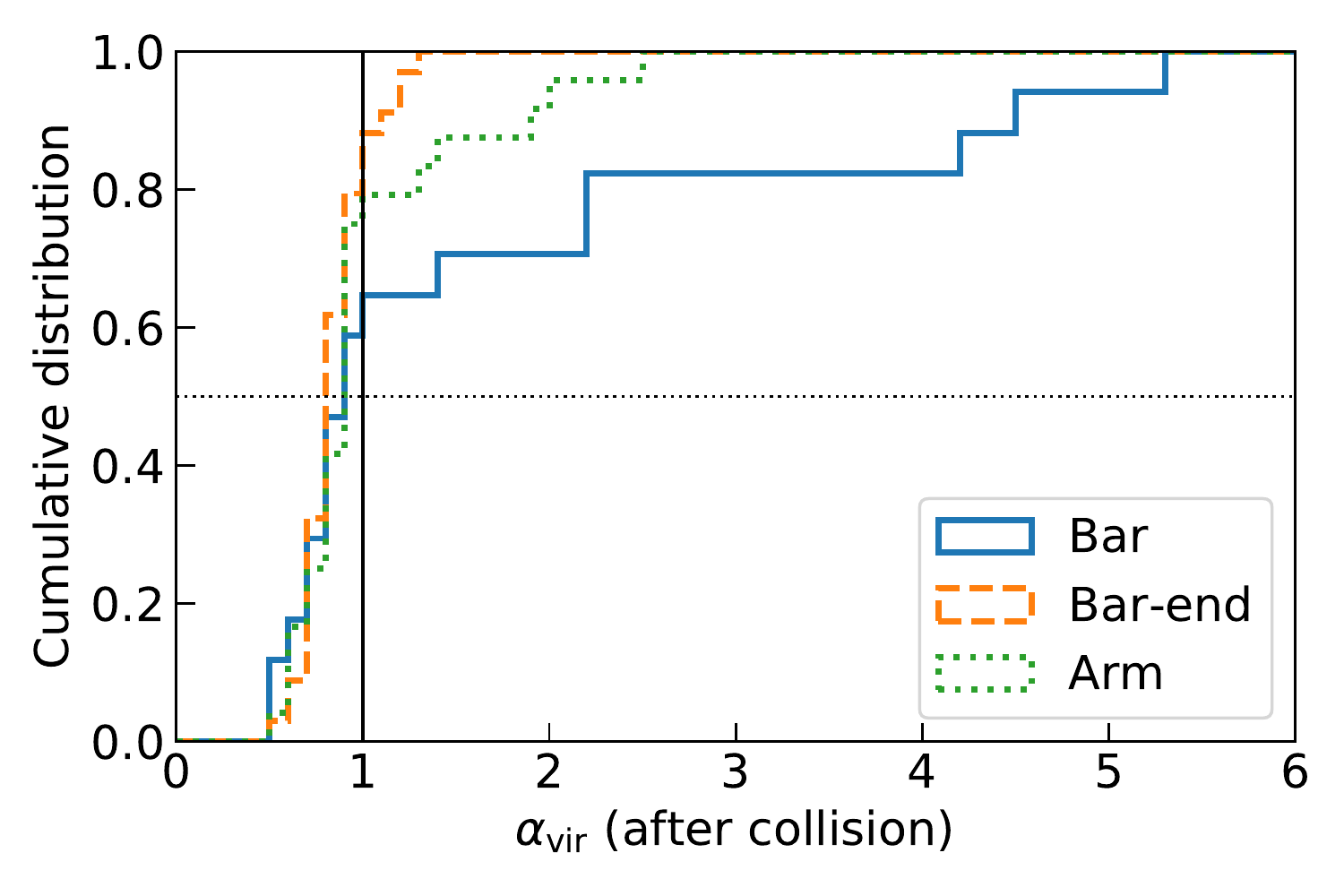}
    \caption{Normalized cumulative distribution function of the virial parameter of clouds after cloud-cloud collisions. The samples are the collisions occurred in 20 Myr between $t = $ 590 and 610 Myr, as shown in Table~\ref{tab:collision_frequencies}. The collision whose mass ratio is less than 0.1 is excluded because we do not expect that such a collision can create a compressed shocked region which could possess massive cloud cores.}
    \label{fig:collision_virial}
\end{figure}

Lastly, in Figure~\ref{fig:collision_virial}, we show the normalized cumulative distribution functions of the virial parameter just after cloud-cloud collision. Even though we stated in Section~\ref{sec:Cloud properties} that the virial parameter might not be an important factor for massive star formation, it is still interesting to check the distribution just after collisions because we can see the effects of the high-speed collision on the cloud's internal state. Figure~\ref{fig:collision_virial} shows that although there is no difference in the median, the \textit{Bar} clouds have a high end tail which is $\alpha_{\mathrm{vir}} > 2$. It indicates that the high-speed collisions might induce turbulence into the cloud, and then parts of the clouds that had experienced the merger could be gravitationally unbound, that is not suitable for formation of low- and intermediate-mass stars, let alone massive stars. 

In summary, we have shown that although there is no significant difference in the cloud's properties such as virial parameter and the frequency of cloud-cloud collision between three galactic regions, the collision speed shows a clear environmental dependence; the \textit{Bar} clouds have $\sim 5\ \mathrm{km\ s^{-1}}$ higher median value than the other regions. 
Significant fraction of clouds collides each other with more than 20 km/s in the bar region. The high-speed collisions in the \textit{Bar} originate from the global galactic gas motion; the gas flow is highly elongated by the bar potential, that induces dispersion in the clouds' motion, that can cause violent cloud-cloud collisions. The high-speed collision can make clouds gravitationally unbound, as we have shown that clouds just after merging can have high virial parameters.

\section{Discussion}
\label{sec:Discussion}

\subsection{The physical mechanism of the fast collision}
\label{sec:The physical mechanism of the fast collision}

Here we make the discussion of the physical mechanism that causes the high-speed cloud collisions in the \textit{Bar} region more quantitatively by examining the spatial distributions of clouds and the velocity deviation between the cloud and surrounding gas, and by estimating the collision frequency of clouds assuming that clouds are in random motion. 

\begin{figure*}
	\includegraphics[width=\columnwidth]{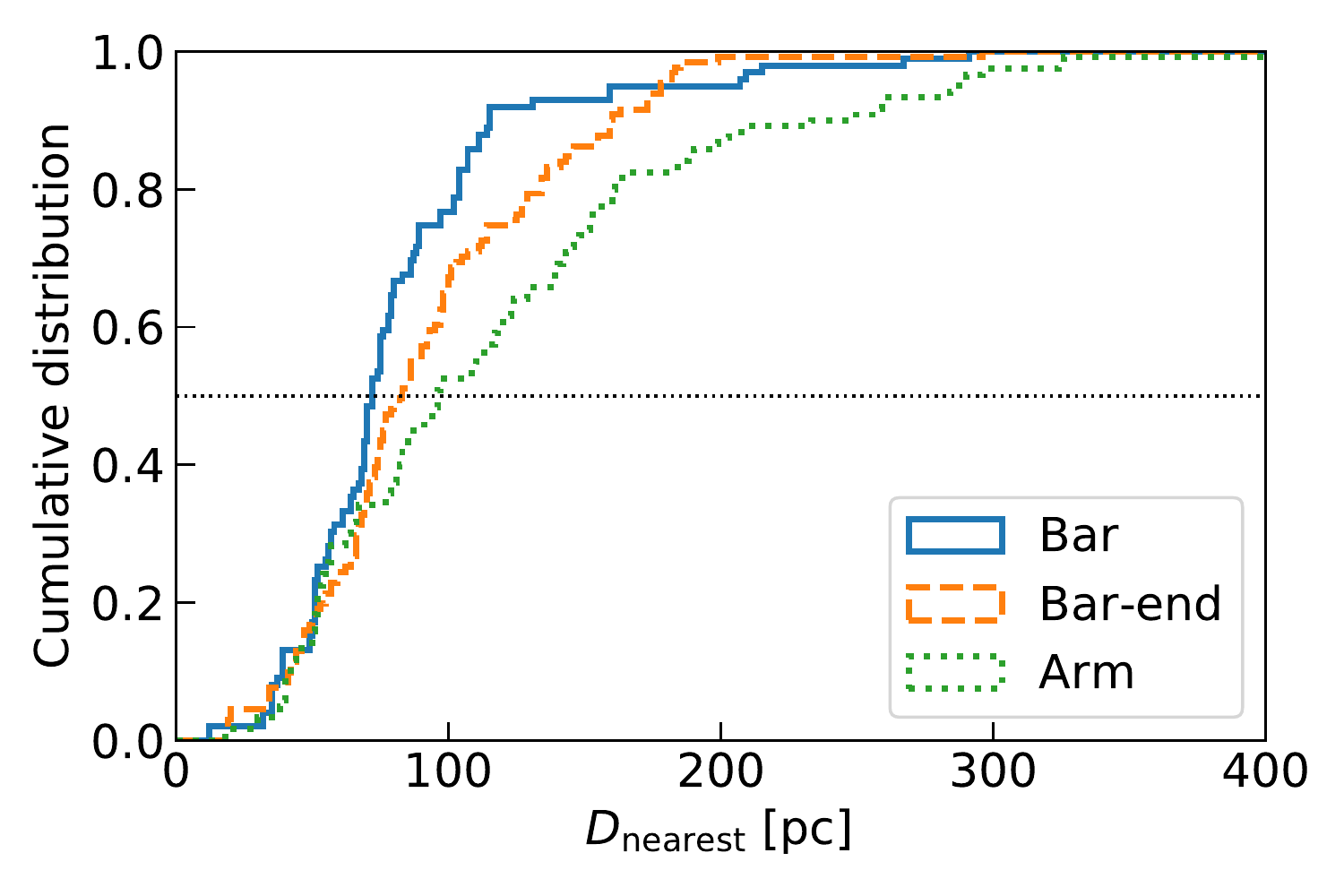}
	\includegraphics[width=\columnwidth]{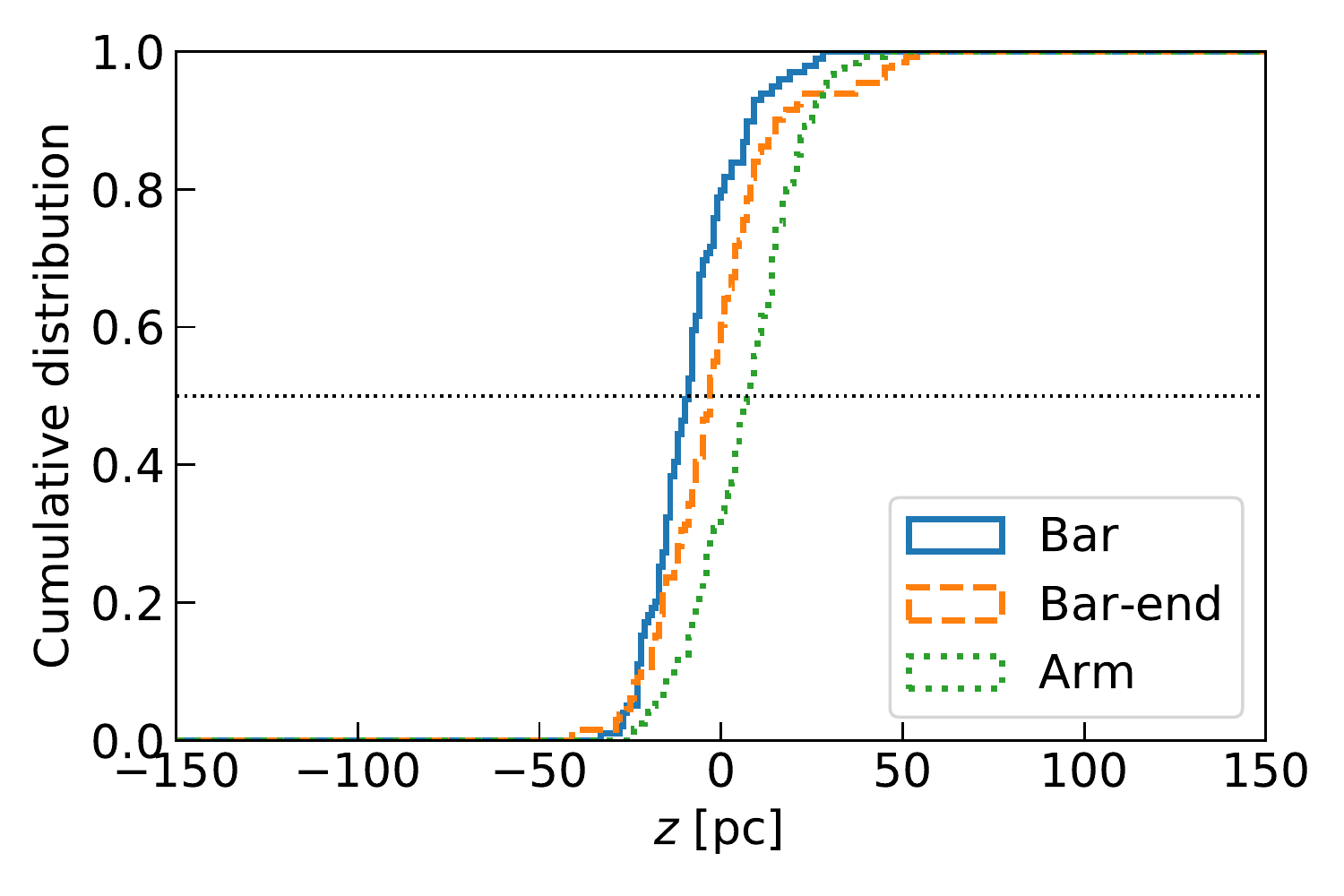}
	\includegraphics[width=\columnwidth]{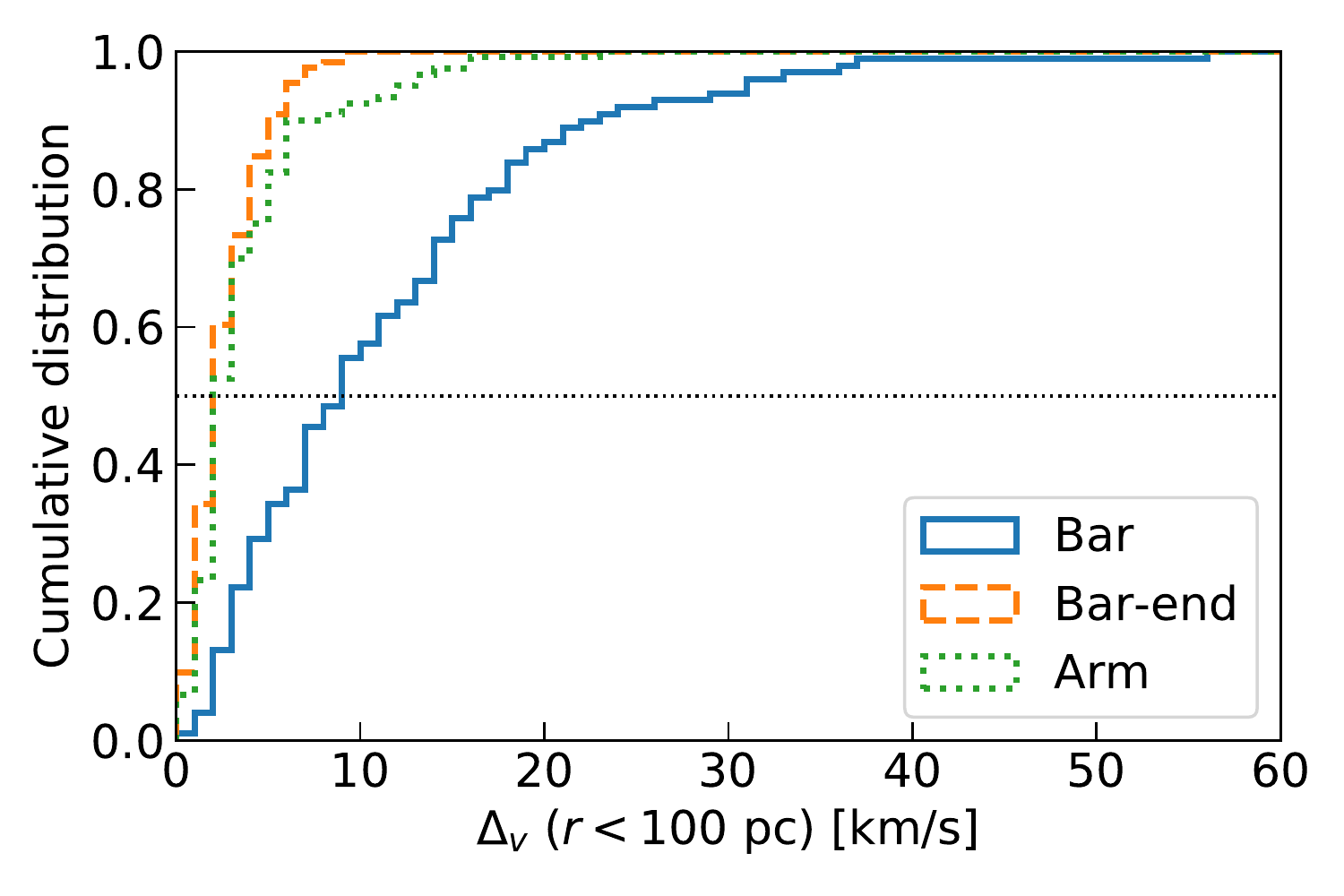}
	\includegraphics[width=\columnwidth]{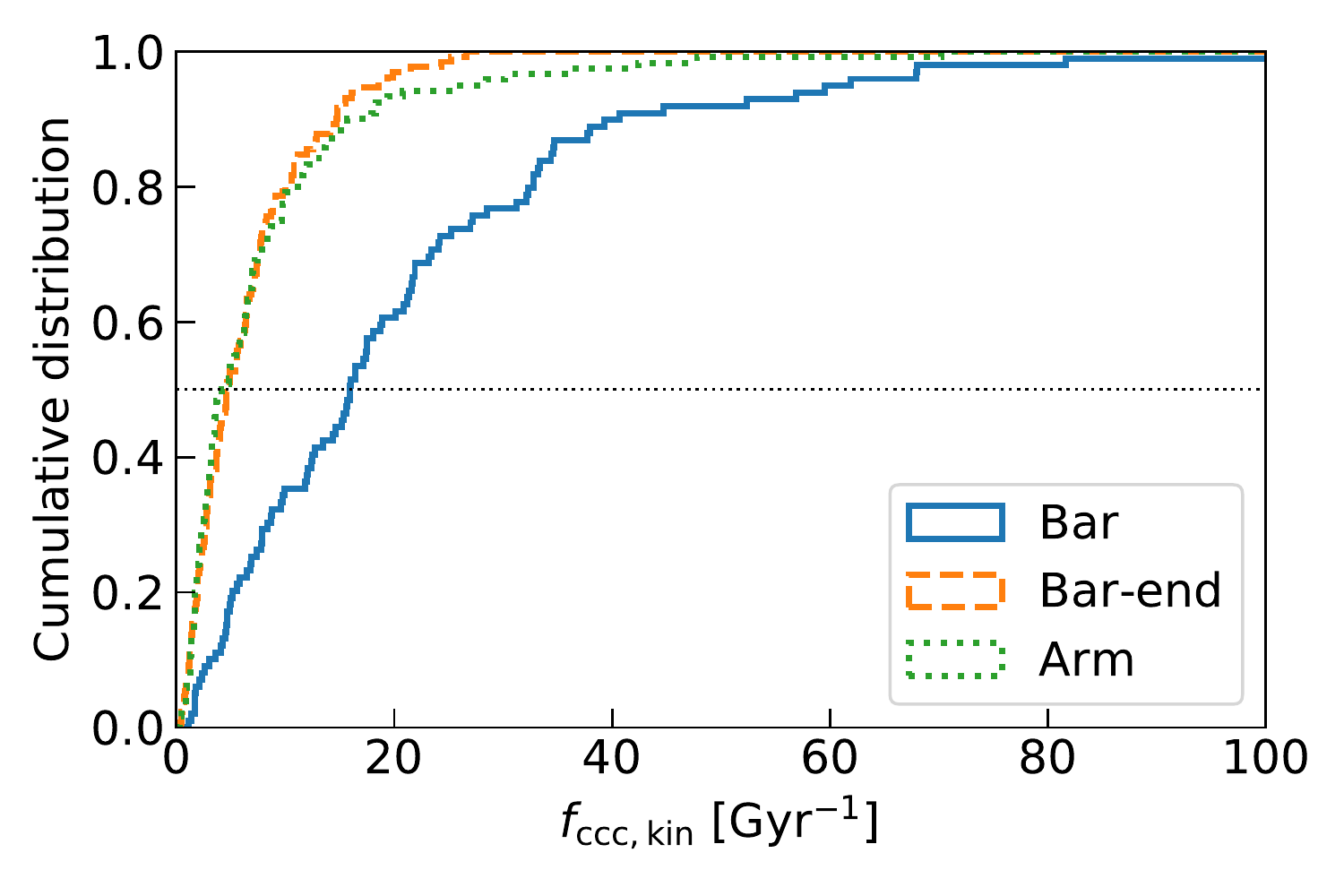}
    \caption{Normalized cumulative distribution function of the distance to the nearest cloud (top left), the vertical cloud distributions (top right), the velocity deviation between the cloud and its surrounding gas within a radius of $r < 100$ pc (bottom left), and the estimated collision frequency of clouds (bottom right).}
    \label{fig:cloud_properties_spatial_distributions}
\end{figure*}

The top left panel of Figure~\ref{fig:cloud_properties_spatial_distributions} shows the normalized cumulative distribution function of the distance to the nearest cloud. Although the difference between regions is only a few tens pc at the median, we see that the \textit{Bar} clouds have a smaller distance to the nearest clouds, indicating that the spatial distribution of the \textit{Bar} clouds are more crowded and clustered compared to those in other two regions. 

The top right panel of Figure~\ref{fig:cloud_properties_spatial_distributions} shows the normalized cumulative distribution function of the vertical cloud distributions. We see that almost all clouds lie in the small region between $-30\ \mathrm{pc} < z < 30\ \mathrm{pc}$ around the galactic mid-plane. It indicates that we can neglect the effect of the vertical cloud distribution in discussion of clouds' motion in $x$-$y$ plane in the following section.

The bottom left panel of Figure~\ref{fig:cloud_properties_spatial_distributions} shows the normalized cumulative distribution function of the velocity deviation between the cloud and its surrounding gas within a radius of $r < 100$ pc. To calculate velocities of the clouds and its surrounding gas, we use a mass-weighted average over the cells for each. We set a cut-off radius of 100 pc because it is nearly the median of the distance to the nearest cloud, as shown in the top left panel of Figure~\ref{fig:cloud_properties_spatial_distributions}.

We find that the \textit{Bar} clouds have a substantially higher velocity deviation than the other two regions; the median in the \textit{Bar} is $\sim 10\ \mathrm{km\ s^{-1}}$, and those in the \textit{Bar-end} and \textit{Arm} are $\sim 2\ \mathrm{km\ s^{-1}}$. The trend of this environmental dependence is similar to that is seen in the cloud-cloud collision speed, as shown in Figure~\ref{fig:collision_speed}. It indicates that the galactic region where clouds have a high-velocity deviation can also have fast cloud-cloud collisions. Note that the collision speeds are faster than the velocity deviations. That might be because, in the collision speed plot, we use only clouds which actually collide with others. On the other hand, in the velocity deviation plot, we use all clouds, including those that have low-velocity deviation and are unlikely to collide soon.

Finally, we estimate the collision frequency of clouds using clouds' kinematics. For this estimate, we impose some assumptions. The first is that all clouds have the same spherical shape and size which is much smaller than the average distance between them. The second is that clouds are in constant, rapid, random motion. The third is that clouds undergo random elastic collisions between themselves. The fourth is that except during collisions, the interactions among clouds such as gravity and viscosity are negligible. This model is well known as kinetic theory and often applied to ideal gases. We also assume that clouds move in a two-dimensional plane as we have shown that we can neglect the vertical cloud distributions in the galactic mid-plane. The collision frequency becomes,
\begin{equation}
    f_{\mathrm{ccc, kin}} = 2 R_{xy} N \Delta_v,
\end{equation}
where $R_{xy} = \sqrt{A_{xy} / \pi}$, $A_{xy}$ is the projected area of the cloud in the $x$-$y$ plane, $N$ is the surface number density of clouds, and $\Delta_v$ is the velocity deviation between the cloud and its surrounding gas as defined above. The $N$ and $\Delta_v$ are calculated within a sphere whose radius is 100 pc centred at the cloud. To calculate frequencies for every clouds, we use $R_{xy}$, $N$, and $\Delta_v$ for each cloud. The bottom right panel of Figure~\ref{fig:cloud_properties_spatial_distributions} shows the normalized cumulative distribution function of the estimated collision frequency of clouds. 

The median frequencies in the \textit{Bar-end} and \textit{Arm} regions are $\sim 5\ \mathrm{Gyr}^{-1}$. They are much lower than the actual frequencies of 17.6 and 13.8 $\mathrm{Gyr}^{-1}$ measured by our cloud tracking as shown in Table~\ref{tab:collision_frequencies}. It is natural to get this result because the \textit{Bar-end} and \textit{Arm} clouds are in gentle motion, and that the gravitational interactions between clouds seem to have a large contribution to collisions as shown in Figure~\ref{fig:collision_scatter_plots}, which is far different from our assumption that clouds are in rapid random motion and that gravitational interactions among them are negligible.

Surprisingly, on the other hand, the clouds' kinematic estimate seems to work in the \textit{Bar}; the median is $\sim 16\ \mathrm{Gyr}^{-1}$, which is almost consistent with the actual collision frequency of $\sim 14.1\ \mathrm{Gyr}^{-1}$, indicating that their clouds' motion has a large deviation so strong as gravitational interactions can be negligible. It suggests that a random-like motion of clouds induced by the elongated gas stream due to the bar potential could be the physical mechanism of the fast collisions in the \textit{Bar} region.

\subsection{Why does not the fast collision affect global distributions of cloud properties?}
\label{Why does not the fast collision affect global distributions of cloud properties?}

\begin{figure*}
	\includegraphics[width=\columnwidth]{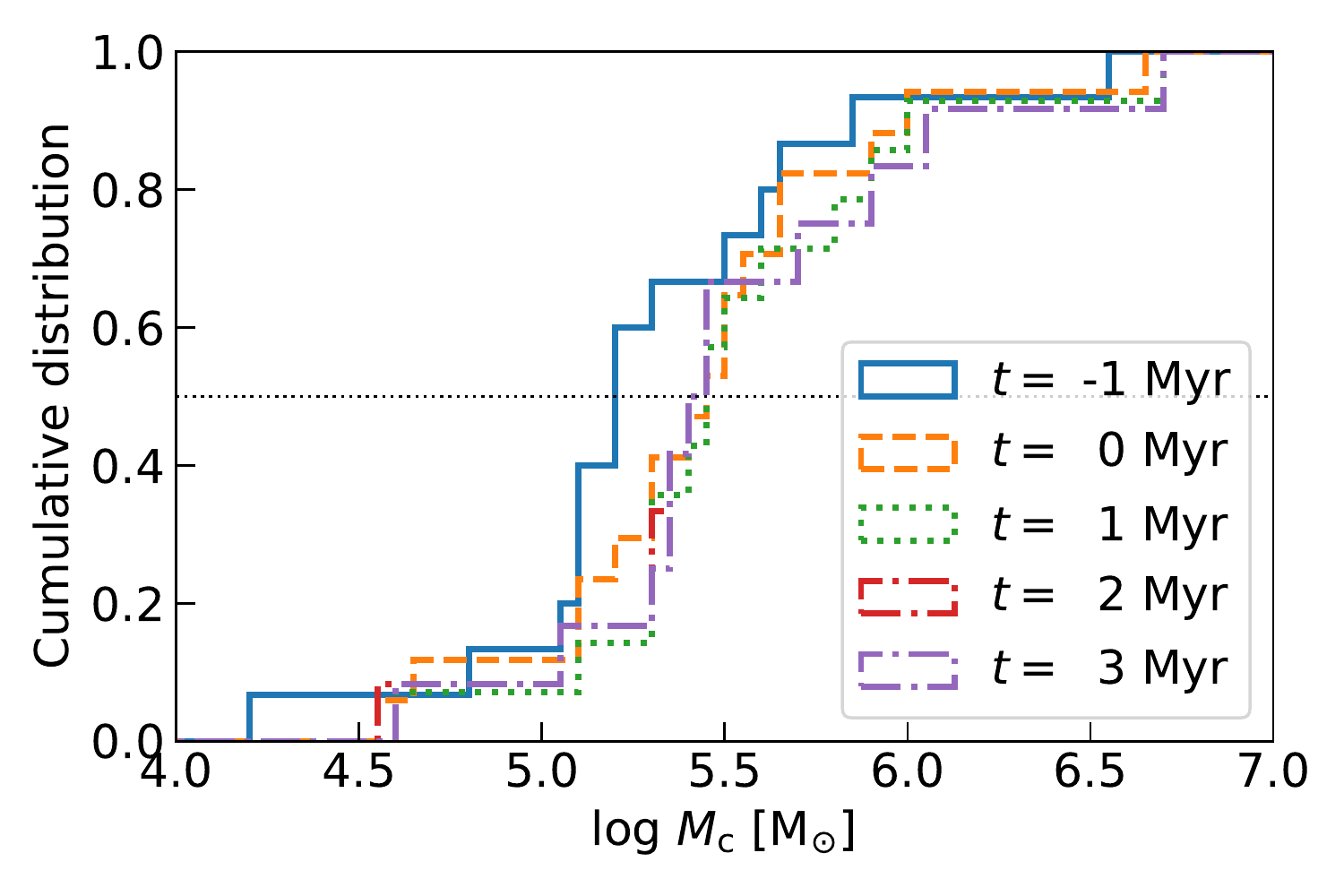}
	\includegraphics[width=\columnwidth]{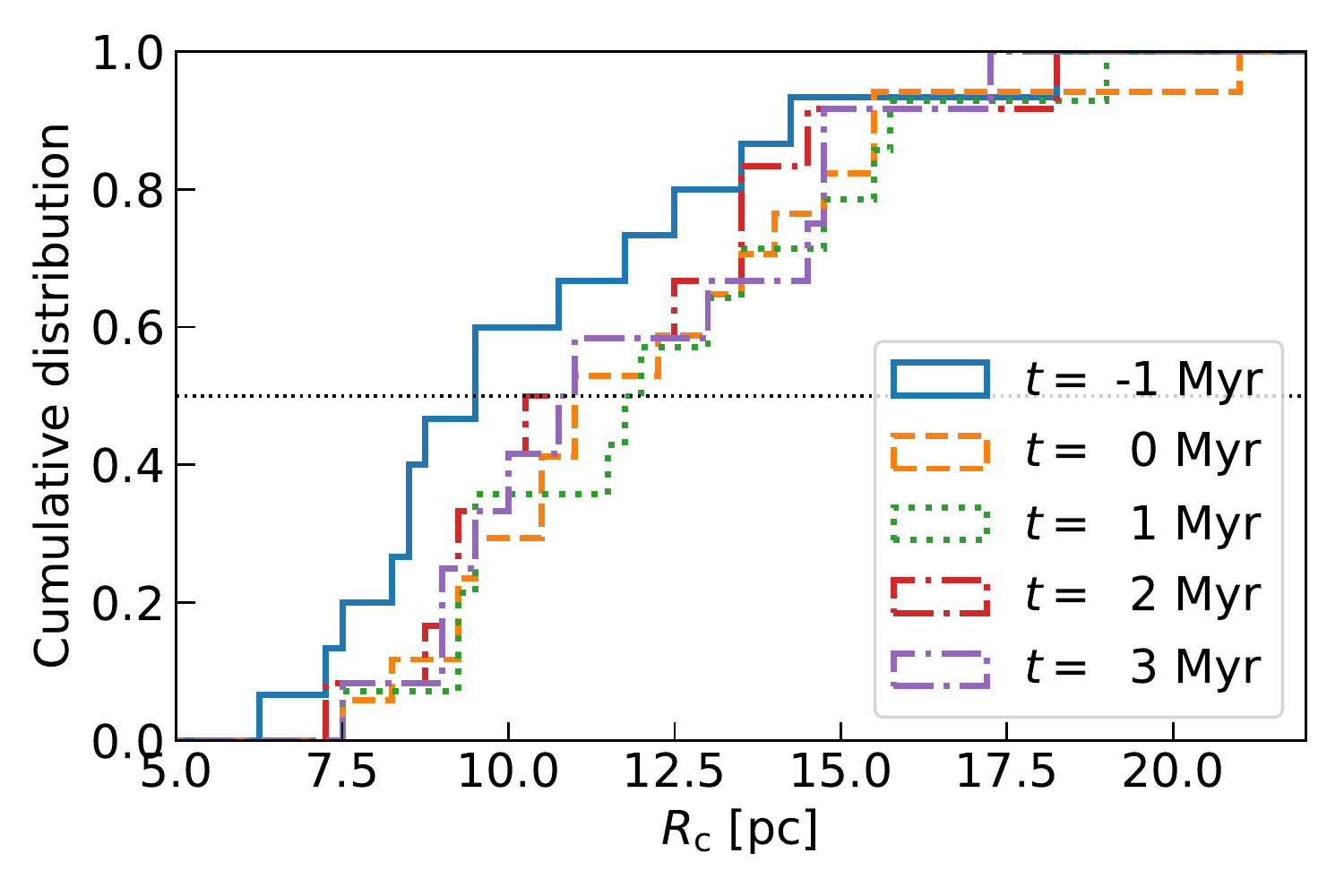}
	\includegraphics[width=\columnwidth]{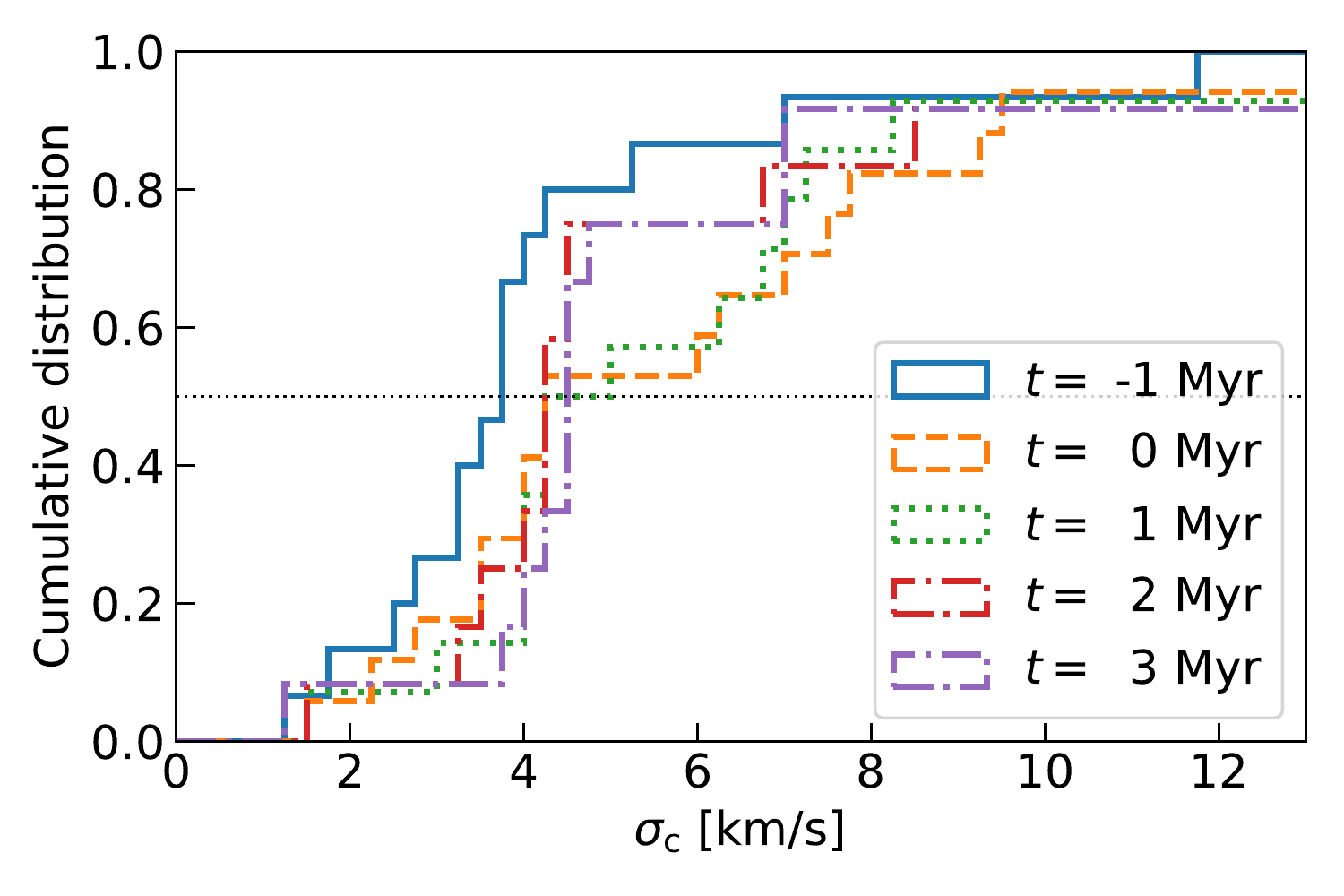}
	\includegraphics[width=\columnwidth]{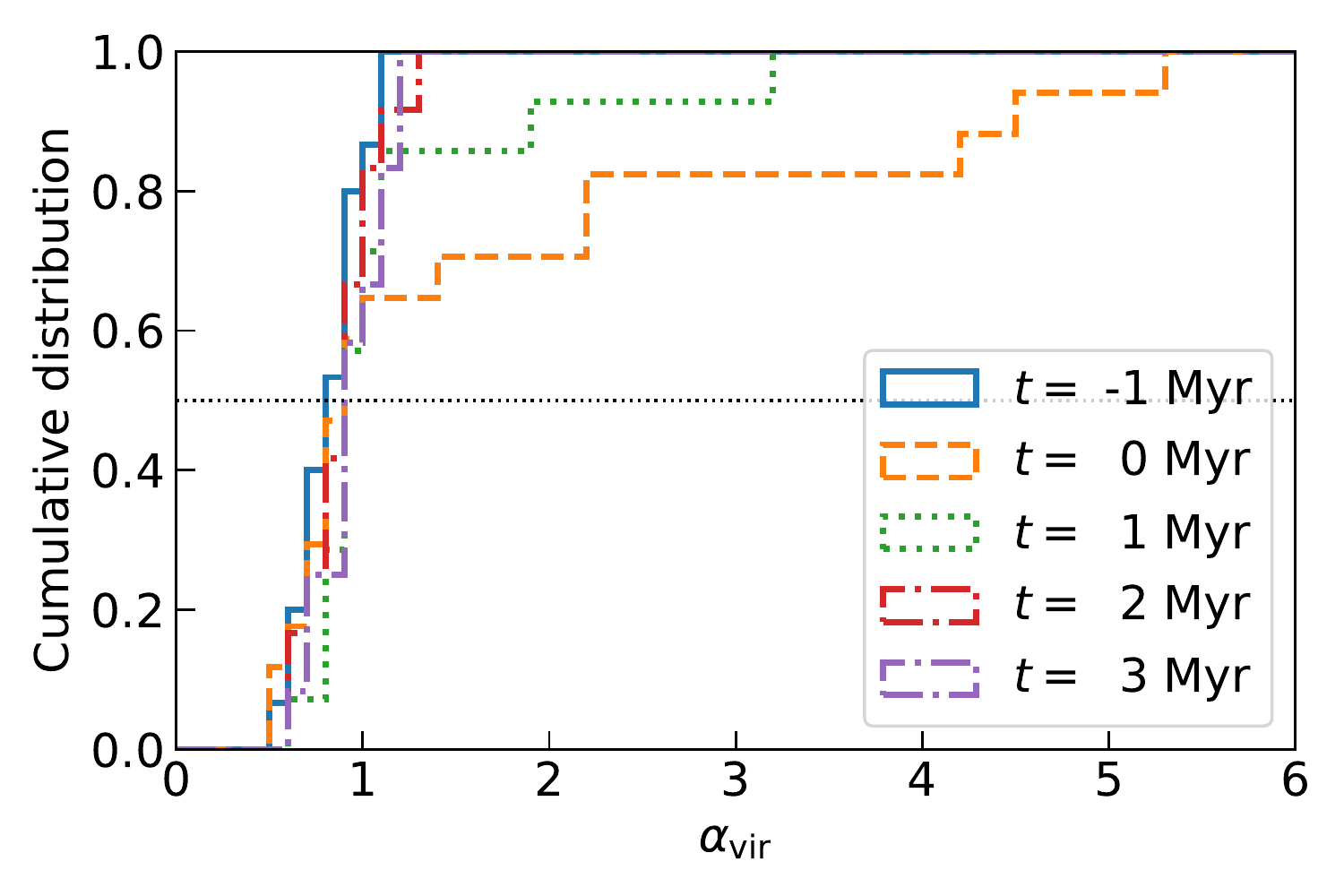}
    \caption{Time evolution of normalized cumulative distribution functions of cloud masses (top left), radii (top right), velocity dispersions (bottom left), and virial parameters (bottom right) before and after cloud collisions in the \textit{Bar} region. The PDFs are at $t = $ -1, 0, 1, 2, and 3 Myr; the time just after collision occurrs is set as $t = $ 0 Myr. Note that there might be a time lag of at most 0.2 Myr because we analyse simulation outputs at intervals of 0.2 Myr.}
    \label{fig:cloud_properties_time_evolution}
\end{figure*}

Here we discuss why the effect of the high-speed collision in the \textit{Bar} region is invisible in the distribution functions of cloud properties; Figure~\ref{fig:cloud_properties} have shown that there is no significant environmental dependence even though the fast cloud collision, which could inject kinetic energy into clouds and make them gravitationally unbound, should be observable in the velocity dispersion and virial parameter. There are two reasons. The first is that the total number of colliding clouds is too small. As shown in Table~\ref{tab:collision_frequencies}, the number of collisions occurred in 20 Myr are 28, 46 and 33 in the \textit{Bar}, \textit{Bar-end} and \textit{Arm} regions, respectively. It indicates that the number of collisions we could find in a one-time snapshot (assuming that the duration is 1 Myr) are only 1.4, 2.3, and 1.65, respectively. These numbers are much smaller than the total number of clouds at $t =$ 600 Myr; the number fractions of the colliding clouds are less than 0.02. 

The second reason is that the effects of the collision do not last for long time; the excited internal gas motion induced by the collision decays quickly within a few Myr. We have investigated the time evolution of cloud properties before and after they collide. Figure~\ref{fig:cloud_properties_time_evolution} shows the cumulative distribution function of the cloud mass, radius, velocity dispersion, and viral parameter at $t = $ -1, 0, 1, 2, and 3 Myr (here, $t = $ 0 Myr is set as the time just after they collide). To see an extreme case, the plots shows only the \textit{Bar} regions where the collision speed is the fastest. Looking at cloud mass and radius distribution, we see that the median values slightly increase just after collision, and they do not get back to the initial value within 3 Myr after the collision. It indicates that the collision forms bigger clouds rather than splitting them into smaller clouds or dispersing the clouds to diffuse ambient gas. Looking at the velocity dispersion, on the other hand, we see a significant increase just after collision, and then they get back to the values close to the initial. The behavior of the viral parameter is similar to the one of the velocity dispersion; 40 per cent of colliding clouds becomes highly unstable after collision, but they return to marginally gravitational bound within at most 2 Myr because of the decay of the excited internal gas velocity dispersion. For those two reasons, it is reasonable that we hardly see the effects of the fast cloud-cloud collisions on global distributions of cloud properties, in particular, the velocity dispersion and virial parameter. 

We emphasize that even though the number fraction of colliding clouds we find in a one-time snapshot looks low, the star formation rate estimated with the assumption that star formation is triggered only by cloud-cloud collisions is roughly consistent with the observations of NGC1300 (except for the \textit{Bar} region, as we discussed in Section~\ref{sec:Cloud-cloud collisions}), indicating that the number of collisions is enough to explain the observed star formation activities.

\subsection{Stellar feedback}
\label{sec:Stellar feedback}

Not only global galactic gas dynamics but also local stellar feedback affect the GMC's formation and evolution \citep[e.g.][]{DobbsBurkertPringle2011, TaskerWadsleyPudritz2015, FujimotoEtAl2016, FujimotoEtAl2019, GrisdaleEtAl2018}. Observations, however, show that the bar structure of a strongly barred galaxy, which is the target of this paper, lacks active star-forming regions, so we expect that the effects of stellar feedback on clouds in the strong bar region are very small and that the formation and evolution scenarios of the \textit{Bar} clouds obtained from our simulation should be more or less consistent with those in the real galaxies. Nonetheless, we can not neglect the local feedback effects, especially in the \textit{Bar-end} and \textit{Arm} regions. In fact, observations of the strongly barred galaxy NGC1300, which is the galaxy modelled by this simulation, show that the bar-end and spiral arm regions contain a lot of H~\textsc{ii} regions and thus have potential to possess massive stars which will explode as supernovae, although the bar region shows almost no star-forming activities. In future work, we will extend our simulations to include stellar feedback in order to have the gas and cloud evolution in the \textit{Bar-end} and \textit{Arm} regions more realistic.

\section{Conclusions}
\label{sec:Conclusions}

Bars of the strongly barred galaxies show an absence of prominent H\textsc{ii} regions even though there are remarkable dust lanes and molecular gas along the stellar bar. The physical mechanism that prevents gas clouds from forming massive stars is still debated. To address this question, we have performed a hydrodynamical simulation of the strongly barred galaxy NGC1300. Our main results are as follows.

\begin{enumerate}
    \item We compare the cumulative distribution functions of cloud properties such as mass, radius, internal velocity dispersion, and virial parameter. We find no significant environmental dependence, in particular, in the virial parameter (Figure~\ref{fig:cloud_properties}). It appears that the observational lack of massive star formation in the strong bar cannot be explained by a systematic trend of clouds which are gravitationally unbound.
    \item Instead, the relative speed of cloud-cloud collisions shows a clear environmental dependence; the \textit{Bar} clouds collide with others faster than those in the other regions (Figure~\ref{fig:collision_speed}). It suggests that the observational lack of massive star formation in the strong bar is caused by fast cloud-cloud collisions, which are inefficient in forming massive stars. The fast collisions result in producing clouds whose virial parameters are extremely high just after the collision ($\alpha_{\mathrm{vir}} > 2$; Figure~\ref{fig:collision_virial}).
    \item The high-speed collisions originate from the galactic-scale elliptical gas motion elongated by the stellar bar potential (Figure~\ref{fig:zoom_projections}). The elongated gas motion induces irregular motion of the ISM and makes the velocity deviations between clouds high in the \textit{Bar} region (Figure~\ref{fig:cloud_properties_spatial_distributions}).
    \item We estimate the collision frequency of clouds using the clouds' kinematics and compare it with the actual frequency measured with cloud tracking. The estimated frequency of the \textit{Bar} cloud is consistent with the actual frequency (Figure~\ref{fig:cloud_properties_spatial_distributions} and Table~\ref{tab:collision_frequencies}), indicating that the high-speed collisions in the \textit{Bar} can be explained by the random-like motion of clouds due to the galactic-scale violent gas motion. 
\end{enumerate}

From these results, we conclude that the physical mechanism that causes the lack of massive star forming regions in the bar of strongly barred galaxies is the high-speed cloud-cloud collisions due to the elongated global gas motion by the stellar bar. For further understanding, we will include stellar feedback to our simulations and compare with high-resolution observations in future work.

\section*{Acknowledgements}

Simulations were carried out on the Cray XC30 and XC50 at the Center for Computational Astrophysics (CfCA) of the National Astronomical Observatory of Japan. YF acknowledges support from the Australian Government through the Australian Research Council's \textit{Discovery Projects} funding scheme (project DP160100695). FM is supported by Research Fellowship for Young Scientists from the JSPS. AH is funded by the JSPS KAKENHI Grant Number JP19K03923. KO is supported by the Grants-in-Aid for Scientific Research (C) (16K05294 and 19K03928) from the Japan Society of the Promotion of Science (JSPS). Computations described in this work were performed using the publicly available \textsc{enzo} code (\citealt{BryanEtAl2014}; \url{http://enzo-project.org}), which is the product of a collaborative effort of many independent scientists from numerous institutions around the world. Their commitment to open science has helped make this work possible. We acknowledge extensive use of the \textsc{yt} package (\citealt{TurkEtAl2011}; \url{http://yt-project.org}) in analysing these results and the authors would like to thank the \textsc{yt} development team for their generous help. This paper makes use of the following ALMA data: ADS/JAO.ALMA\#2015.1.00925.S. ALMA is a partnership of ESO (representing its member states), NSF (USA) and NINS (Japan), together with NRC (Canada), MOST and ASIAA (Taiwan), and KASI (Republic of Korea), in cooperation with the Republic of Chile. The Joint ALMA Observatory is operated by ESO, AUI/NRAO, and NAOJ.




\bibliographystyle{mnras}
\bibliography{reference} 

\begin{thebibliography}{}
\makeatletter
\relax
\def\mn@urlcharsother{\let\do\@makeother \do\$\do\&\do\#\do\^\do\_\do\%\do\~}
\def\mn@doi{\begingroup\mn@urlcharsother \@ifnextchar [ {\mn@doi@}
  {\mn@doi@[]}}
\def\mn@doi@[#1]#2{\def\@tempa{#1}\ifx\@tempa\@empty \href
  {http://dx.doi.org/#2} {doi:#2}\else \href {http://dx.doi.org/#2} {#1}\fi
  \endgroup}
\def\mn@eprint#1#2{\mn@eprint@#1:#2::\@nil}
\def\mn@eprint@arXiv#1{\href {http://arxiv.org/abs/#1} {{\tt arXiv:#1}}}
\def\mn@eprint@dblp#1{\href {http://dblp.uni-trier.de/rec/bibtex/#1.xml}
  {dblp:#1}}
\def\mn@eprint@#1:#2:#3:#4\@nil{\def\@tempa {#1}\def\@tempb {#2}\def\@tempc
  {#3}\ifx \@tempc \@empty \let \@tempc \@tempb \let \@tempb \@tempa \fi \ifx
  \@tempb \@empty \def\@tempb {arXiv}\fi \@ifundefined
  {mn@eprint@\@tempb}{\@tempb:\@tempc}{\expandafter \expandafter \csname
  mn@eprint@\@tempb\endcsname \expandafter{\@tempc}}}

\bibitem[\protect\citeauthoryear{{Abel}, {Anninos}, {Zhang}  \&
  {Norman}}{{Abel} et~al.}{1997}]{AbelEtAl1997}
{Abel} T.,  {Anninos} P.,  {Zhang} Y.,   {Norman} M.~L.,  1997, \mn@doi [New
  Astronomy] {10.1016/S1384-1076(97)00010-9}, \href
  {https://ui.adsabs.harvard.edu/abs/1997NewA....2..181A} {2, 181}

\bibitem[\protect\citeauthoryear{{Athanassoula}}{{Athanassoula}}{1992}]{Athanassoula1992}
{Athanassoula} E.,  1992, \mn@doi [\mnras] {10.1093/mnras/259.2.345}, \href
  {https://ui.adsabs.harvard.edu/abs/1992MNRAS.259..345A} {259, 345}

\bibitem[\protect\citeauthoryear{{Bertoldi} \& {McKee}}{{Bertoldi} \&
  {McKee}}{1992}]{BertoldiMcKee1992}
{Bertoldi} F.,  {McKee} C.~F.,  1992, \mn@doi [\apj] {10.1086/171638}, \href
  {https://ui.adsabs.harvard.edu/abs/1992ApJ...395..140B} {395, 140}

\bibitem[\protect\citeauthoryear{{Bolatto}, {Wolfire}  \& {Leroy}}{{Bolatto}
  et~al.}{2013}]{BolattoWolfireLeroy2013}
{Bolatto} A.~D.,  {Wolfire} M.,   {Leroy} A.~K.,  2013, \mn@doi [\araa]
  {10.1146/annurev-astro-082812-140944}, \href
  {https://ui.adsabs.harvard.edu/abs/2013ARA&A..51..207B} {51, 207}

\bibitem[\protect\citeauthoryear{{Brummel-Smith} et~al.,}{{Brummel-Smith}
  et~al.}{2019}]{BrummelEtAl2019}
{Brummel-Smith} C.,  et~al., 2019, \mn@doi [The Journal of Open Source
  Software] {10.21105/joss.01636}, \href
  {https://ui.adsabs.harvard.edu/abs/2019JOSS....4.1636B} {4, 1636}

\bibitem[\protect\citeauthoryear{{Bryan} et~al.,}{{Bryan}
  et~al.}{2014}]{BryanEtAl2014}
{Bryan} G.~L.,  et~al., 2014, \mn@doi [\apjs] {10.1088/0067-0049/211/2/19},
  \href {http://adsabs.harvard.edu/abs/2014ApJS..211...19B} {211, 19}

\bibitem[\protect\citeauthoryear{{Chevance} et~al.,}{{Chevance}
  et~al.}{2019}]{ChevanceEtAl2019}
{Chevance} M.,  et~al., 2019, \mn@doi [\mnras] {10.1093/mnras/stz3525}, \href
  {https://ui.adsabs.harvard.edu/abs/2019MNRAS.tmp.3155C} {p.~3155}

\bibitem[\protect\citeauthoryear{{Daddi} et~al.,}{{Daddi}
  et~al.}{2010}]{DaddiEtAl2010}
{Daddi} E.,  et~al., 2010, \mn@doi [\apj] {10.1088/2041-8205/714/1/L118}, \href
  {https://ui.adsabs.harvard.edu/abs/2010ApJ...714L.118D} {714, L118}

\bibitem[\protect\citeauthoryear{{Dobbs}, {Burkert}  \& {Pringle}}{{Dobbs}
  et~al.}{2011}]{DobbsBurkertPringle2011}
{Dobbs} C.~L.,  {Burkert} A.,   {Pringle} J.~E.,  2011, \mn@doi [\mnras]
  {10.1111/j.1365-2966.2011.19346.x}, \href
  {https://ui.adsabs.harvard.edu/abs/2011MNRAS.417.1318D} {417, 1318}

\bibitem[\protect\citeauthoryear{{Dobbs}, {Pringle}  \&
  {Duarte-Cabral}}{{Dobbs} et~al.}{2015}]{DobbsPringleDuarte-Cabral2015}
{Dobbs} C.~L.,  {Pringle} J.~E.,   {Duarte-Cabral} A.,  2015, \mn@doi [\mnras]
  {10.1093/mnras/stu2319}, \href
  {https://ui.adsabs.harvard.edu/abs/2015MNRAS.446.3608D} {446, 3608}

\bibitem[\protect\citeauthoryear{{Downes}, {Reynaud}, {Solomon}  \&
  {Radford}}{{Downes} et~al.}{1996}]{DownesEtAl1996}
{Downes} D.,  {Reynaud} D.,  {Solomon} P.~M.,   {Radford} S.~J.~E.,  1996,
  \mn@doi [\apj] {10.1086/177046}, \href
  {https://ui.adsabs.harvard.edu/abs/1996ApJ...461..186D} {461, 186}

\bibitem[\protect\citeauthoryear{{Draine}}{{Draine}}{2011}]{Draine2011}
{Draine} B.~T.,  2011, {Physics of the Interstellar and Intergalactic Medium}.
piim.book

\bibitem[\protect\citeauthoryear{{England}}{{England}}{1989a}]{England1989a}
{England} M.~N.,  1989a, \mn@doi [\apj] {10.1086/167097}, \href
  {http://adsabs.harvard.edu/abs/1989ApJ...337..191E} {337, 191}

\bibitem[\protect\citeauthoryear{{England}}{{England}}{1989b}]{England1989b}
{England} M.~N.,  1989b, \mn@doi [\apj] {10.1086/167833}, \href
  {http://ads.nao.ac.jp/abs/1989ApJ...344..669E} {344, 669}

\bibitem[\protect\citeauthoryear{{Ferland}, {Korista}, {Verner}, {Ferguson},
  {Kingdon}  \& {Verner}}{{Ferland} et~al.}{1998}]{FerlandEtAl1998}
{Ferland} G.~J.,  {Korista} K.~T.,  {Verner} D.~A.,  {Ferguson} J.~W.,
  {Kingdon} J.~B.,   {Verner} E.~M.,  1998, \mn@doi [\pasp] {10.1086/316190},
  \href {https://ui.adsabs.harvard.edu/abs/1998PASP..110..761F} {110, 761}

\bibitem[\protect\citeauthoryear{{Freeman}, {Rosolowsky}, {Kruijssen},
  {Bastian}  \& {Adamo}}{{Freeman} et~al.}{2017}]{FreemanEtAl2017}
{Freeman} P.,  {Rosolowsky} E.,  {Kruijssen} J.~M.~D.,  {Bastian} N.,   {Adamo}
  A.,  2017, \mn@doi [\mnras] {10.1093/mnras/stx499}, \href
  {http://adsabs.harvard.edu/abs/2017MNRAS.468.1769F} {468, 1769}

\bibitem[\protect\citeauthoryear{{Fujimoto}, {Tasker}, {Wakayama}  \&
  {Habe}}{{Fujimoto} et~al.}{2014a}]{FujimotoEtAl2014}
{Fujimoto} Y.,  {Tasker} E.~J.,  {Wakayama} M.,   {Habe} A.,  2014a, \mn@doi
  [\mnras] {10.1093/mnras/stu014}, \href
  {http://adsabs.harvard.edu/abs/2014MNRAS.439..936F} {439, 936}

\bibitem[\protect\citeauthoryear{{Fujimoto}, {Tasker}  \& {Habe}}{{Fujimoto}
  et~al.}{2014b}]{FujimotoTaskerHabe2014}
{Fujimoto} Y.,  {Tasker} E.~J.,   {Habe} A.,  2014b, \mn@doi [\mnras]
  {10.1093/mnrasl/slu138}, \href
  {http://adsabs.harvard.edu/abs/2014MNRAS.445L..65F} {445, L65}

\bibitem[\protect\citeauthoryear{{Fujimoto}, {Bryan}, {Tasker}, {Habe}  \&
  {Simpson}}{{Fujimoto} et~al.}{2016}]{FujimotoEtAl2016}
{Fujimoto} Y.,  {Bryan} G.~L.,  {Tasker} E.~J.,  {Habe} A.,   {Simpson} C.~M.,
  2016, \mn@doi [\mnras] {10.1093/mnras/stw1461}, \href
  {http://adsabs.harvard.edu/abs/2016MNRAS.461.1684F} {461, 1684}

\bibitem[\protect\citeauthoryear{{Fujimoto}, {Chevance}, {Haydon}, {Krumholz}
  \& {Kruijssen}}{{Fujimoto} et~al.}{2019}]{FujimotoEtAl2019}
{Fujimoto} Y.,  {Chevance} M.,  {Haydon} D.~T.,  {Krumholz} M.~R.,
  {Kruijssen} J.~M.~D.,  2019, \mn@doi [\mnras] {10.1093/mnras/stz641}, \href
  {https://ui.adsabs.harvard.edu/abs/2019MNRAS.tmp..625F} {p.~625}

\bibitem[\protect\citeauthoryear{{Fukui} et~al.,}{{Fukui}
  et~al.}{2014}]{FukuiEtAl2014}
{Fukui} Y.,  et~al., 2014, \mn@doi [\apj] {10.1088/0004-637X/780/1/36}, \href
  {https://ui.adsabs.harvard.edu/abs/2014ApJ...780...36F} {780, 36}

\bibitem[\protect\citeauthoryear{{Fukui} et~al.,}{{Fukui}
  et~al.}{2016}]{2016ApJ...820...26F}
{Fukui} Y.,  et~al., 2016, \mn@doi [\apj] {10.3847/0004-637X/820/1/26}, \href
  {https://ui.adsabs.harvard.edu/abs/2016ApJ...820...26F} {820, 26}

\bibitem[\protect\citeauthoryear{{Fukui} et~al.,}{{Fukui}
  et~al.}{2018}]{2018ApJ...859..166F}
{Fukui} Y.,  et~al., 2018, \mn@doi [\apj] {10.3847/1538-4357/aac217}, \href
  {http://ads.nao.ac.jp/abs/2018ApJ...859..166F} {859, 166}

\bibitem[\protect\citeauthoryear{{Furukawa}, {Dawson}, {Ohama}, {Kawamura},
  {Mizuno}, {Onishi}  \& {Fukui}}{{Furukawa} et~al.}{2009}]{FurukawaEtAl2009}
{Furukawa} N.,  {Dawson} J.~R.,  {Ohama} A.,  {Kawamura} A.,  {Mizuno} N.,
  {Onishi} T.,   {Fukui} Y.,  2009, \mn@doi [\apj]
  {10.1088/0004-637X/696/2/L115}, \href
  {https://ui.adsabs.harvard.edu/abs/2009ApJ...696L.115F} {696, L115}

\bibitem[\protect\citeauthoryear{{Grisdale}, {Agertz}, {Renaud}  \&
  {Romeo}}{{Grisdale} et~al.}{2018}]{GrisdaleEtAl2018}
{Grisdale} K.,  {Agertz} O.,  {Renaud} F.,   {Romeo} A.~B.,  2018, \mn@doi
  [\mnras] {10.1093/mnras/sty1595}, \href
  {https://ui.adsabs.harvard.edu/abs/2018MNRAS.479.3167G} {479, 3167}

\bibitem[\protect\citeauthoryear{{Habe} \& {Ohta}}{{Habe} \&
  {Ohta}}{1992}]{HabeOhta1992}
{Habe} A.,  {Ohta} K.,  1992, \pasj, \href
  {https://ui.adsabs.harvard.edu/abs/1992PASJ...44..203H} {44, 203}

\bibitem[\protect\citeauthoryear{{Hirota} et~al.,}{{Hirota}
  et~al.}{2014}]{HirotaEtAl2014}
{Hirota} A.,  et~al., 2014, \mn@doi [\pasj] {10.1093/pasj/psu006}, \href
  {https://ui.adsabs.harvard.edu/abs/2014PASJ...66...46H} {66, 46}

\bibitem[\protect\citeauthoryear{{Jin}, {Salim}, {Federrath}, {Tasker}, {Habe}
  \& {Kainulainen}}{{Jin} et~al.}{2017}]{JinEtAl2017}
{Jin} K.,  {Salim} D.~M.,  {Federrath} C.,  {Tasker} E.~J.,  {Habe} A.,
  {Kainulainen} J.~T.,  2017, \mn@doi [\mnras] {10.1093/mnras/stx737}, \href
  {https://ui.adsabs.harvard.edu/abs/2017MNRAS.469..383J} {469, 383}

\bibitem[\protect\citeauthoryear{{Kawamura} et~al.,}{{Kawamura}
  et~al.}{2009}]{KawamuraEtAl2009}
{Kawamura} A.,  et~al., 2009, \mn@doi [\apjs] {10.1088/0067-0049/184/1/1},
  \href {https://ui.adsabs.harvard.edu/abs/2009ApJS..184....1K} {184, 1}

\bibitem[\protect\citeauthoryear{{Leroy} et~al.,}{{Leroy}
  et~al.}{2009}]{LeroyEtAl2009}
{Leroy} A.~K.,  et~al., 2009, \mn@doi [\aj] {10.1088/0004-6256/137/6/4670},
  \href {https://ui.adsabs.harvard.edu/abs/2009AJ....137.4670L} {137, 4670}

\bibitem[\protect\citeauthoryear{{Leroy} et~al.,}{{Leroy}
  et~al.}{2013}]{LeroyEtAl2013}
{Leroy} A.~K.,  et~al., 2013, \mn@doi [\aj] {10.1088/0004-6256/146/2/19}, \href
  {https://ui.adsabs.harvard.edu/abs/2013AJ....146...19L} {146, 19}

\bibitem[\protect\citeauthoryear{{Li}, {Tan}, {Christie}, {Bisbas}  \&
  {Wu}}{{Li} et~al.}{2018}]{LiEtAl2018}
{Li} Q.,  {Tan} J.~C.,  {Christie} D.,  {Bisbas} T.~G.,   {Wu} B.,  2018,
  \mn@doi [\pasj] {10.1093/pasj/psx136}, \href
  {http://adsabs.harvard.edu/abs/2018PASJ...70S..56L} {70, S56}

\bibitem[\protect\citeauthoryear{{Maeda}, {Ohta}, {Fujimoto}, {Habe}  \&
  {Baba}}{{Maeda} et~al.}{2018}]{MaedaEtAl2018}
{Maeda} F.,  {Ohta} K.,  {Fujimoto} Y.,  {Habe} A.,   {Baba} J.,  2018, \mn@doi
  [\pasj] {10.1093/pasj/psy028}, \href
  {https://ui.adsabs.harvard.edu/abs/2018PASJ...70...37M} {70, 37}

\bibitem[\protect\citeauthoryear{{Maeda}, {Ohta}, {Fujimoto}  \&
  {Habe}}{{Maeda} et~al.}{2020}]{MaedaEtAl2020}
{Maeda} F.,  {Ohta} K.,  {Fujimoto} Y.,   {Habe} A.,  2020, arXiv e-prints,
  \href {https://ui.adsabs.harvard.edu/abs/2020arXiv200208977M} {p.
  arXiv:2002.08977}

\bibitem[\protect\citeauthoryear{{Meidt} et~al.,}{{Meidt}
  et~al.}{2013}]{MeidtEtAl2013}
{Meidt} S.~E.,  et~al., 2013, \mn@doi [\apj] {10.1088/0004-637X/779/1/45},
  \href {https://ui.adsabs.harvard.edu/abs/2013ApJ...779...45M} {779, 45}

\bibitem[\protect\citeauthoryear{{Momose}, {Okumura}, {Koda}  \&
  {Sawada}}{{Momose} et~al.}{2010}]{MomoseEtAl2010}
{Momose} R.,  {Okumura} S.~K.,  {Koda} J.,   {Sawada} T.,  2010, \mn@doi [\apj]
  {10.1088/0004-637X/721/1/383}, \href
  {https://ui.adsabs.harvard.edu/abs/2010ApJ...721..383M} {721, 383}

\bibitem[\protect\citeauthoryear{{Navarro}, {Frenk}  \& {White}}{{Navarro}
  et~al.}{1996}]{NavarroEtAl1996}
{Navarro} J.~F.,  {Frenk} C.~S.,   {White} S.~D.~M.,  1996, \mn@doi [\apj]
  {10.1086/177173}, \href {http://ads.nao.ac.jp/abs/1996ApJ...462..563N} {462,
  563}

\bibitem[\protect\citeauthoryear{{Nimori}, {Habe}, {Sorai}, {Watanabe},
  {Hirota}  \& {Namekata}}{{Nimori} et~al.}{2013}]{NimoriEtAl2013}
{Nimori} M.,  {Habe} A.,  {Sorai} K.,  {Watanabe} Y.,  {Hirota} A.,
  {Namekata} D.,  2013, \mn@doi [\mnras] {10.1093/mnras/sts487}, \href
  {https://ui.adsabs.harvard.edu/abs/2013MNRAS.429.2175N} {429, 2175}

\bibitem[\protect\citeauthoryear{{Ohama} et~al.,}{{Ohama}
  et~al.}{2010}]{OhamaEtAl2010}
{Ohama} A.,  et~al., 2010, \mn@doi [\apj] {10.1088/0004-637X/709/2/975}, \href
  {https://ui.adsabs.harvard.edu/abs/2010ApJ...709..975O} {709, 975}

\bibitem[\protect\citeauthoryear{{Oka}, {Hasegawa}, {Sato}, {Tsuboi},
  {Miyazaki}  \& {Sugimoto}}{{Oka} et~al.}{2001}]{OkaEtAl2001}
{Oka} T.,  {Hasegawa} T.,  {Sato} F.,  {Tsuboi} M.,  {Miyazaki} A.,
  {Sugimoto} M.,  2001, \mn@doi [\apj] {10.1086/322976}, \href
  {https://ui.adsabs.harvard.edu/abs/2001ApJ...562..348O} {562, 348}

\bibitem[\protect\citeauthoryear{{Reynaud} \& {Downes}}{{Reynaud} \&
  {Downes}}{1998}]{ReynaudDownes1998}
{Reynaud} D.,  {Downes} D.,  1998, \aap, \href
  {https://ui.adsabs.harvard.edu/abs/1998A&A...337..671R} {337, 671}

\bibitem[\protect\citeauthoryear{{Rosolowsky}, {Engargiola}, {Plambeck}  \&
  {Blitz}}{{Rosolowsky} et~al.}{2003}]{RosolowskyEtAl2003}
{Rosolowsky} E.,  {Engargiola} G.,  {Plambeck} R.,   {Blitz} L.,  2003, \mn@doi
  [\apj] {10.1086/379166}, \href
  {http://adsabs.harvard.edu/abs/2003ApJ...599..258R} {599, 258}

\bibitem[\protect\citeauthoryear{{Sheth}, {Vogel}, {Regan}, {Teuben}, {Harris}
  \& {Thornley}}{{Sheth} et~al.}{2002}]{ShethEtAl2002}
{Sheth} K.,  {Vogel} S.~N.,  {Regan} M.~W.,  {Teuben} P.~J.,  {Harris} A.~I.,
  {Thornley} M.~D.,  2002, \mn@doi [\aj] {10.1086/343835}, \href
  {https://ui.adsabs.harvard.edu/abs/2002AJ....124.2581S} {124, 2581}

\bibitem[\protect\citeauthoryear{{Sorai} et~al.,}{{Sorai}
  et~al.}{2012}]{SoraiEtAl2012}
{Sorai} K.,  et~al., 2012, \mn@doi [\pasj] {10.1093/pasj/64.3.51}, \href
  {https://ui.adsabs.harvard.edu/abs/2012PASJ...64...51S} {64, 51}

\bibitem[\protect\citeauthoryear{{Stone} \& {Norman}}{{Stone} \&
  {Norman}}{1992}]{StoneNorman1992}
{Stone} J.~M.,  {Norman} M.~L.,  1992, \mn@doi [\apjs] {10.1086/191680}, \href
  {https://ui.adsabs.harvard.edu/abs/1992ApJS...80..753S} {80, 753}

\bibitem[\protect\citeauthoryear{{Takahira}, {Tasker}  \& {Habe}}{{Takahira}
  et~al.}{2014}]{TakahiraTaskerHabe2014}
{Takahira} K.,  {Tasker} E.~J.,   {Habe} A.,  2014, \mn@doi [\apj]
  {10.1088/0004-637X/792/1/63}, \href
  {https://ui.adsabs.harvard.edu/abs/2014ApJ...792...63T} {792, 63}

\bibitem[\protect\citeauthoryear{{Takahira}, {Shima}, {Habe}  \&
  {Tasker}}{{Takahira} et~al.}{2018}]{TakahiraEtAl2018}
{Takahira} K.,  {Shima} K.,  {Habe} A.,   {Tasker} E.~J.,  2018, \mn@doi
  [\pasj] {10.1093/pasj/psy011}, \href
  {https://ui.adsabs.harvard.edu/abs/2018PASJ...70S..58T} {70, S58}

\bibitem[\protect\citeauthoryear{{Tan}}{{Tan}}{2000}]{Tan2000}
{Tan} J.~C.,  2000, \mn@doi [\apj] {10.1086/308905}, \href
  {https://ui.adsabs.harvard.edu/abs/2000ApJ...536..173T} {536, 173}

\bibitem[\protect\citeauthoryear{{Tasker} \& {Tan}}{{Tasker} \&
  {Tan}}{2009}]{TaskerTan2009}
{Tasker} E.~J.,  {Tan} J.~C.,  2009, \mn@doi [\apj]
  {10.1088/0004-637X/700/1/358}, \href
  {http://adsabs.harvard.edu/abs/2009ApJ...700..358T} {700, 358}

\bibitem[\protect\citeauthoryear{{Tasker}, {Wadsley}  \& {Pudritz}}{{Tasker}
  et~al.}{2015}]{TaskerWadsleyPudritz2015}
{Tasker} E.~J.,  {Wadsley} J.,   {Pudritz} R.,  2015, \mn@doi [\apj]
  {10.1088/0004-637X/801/1/33}, \href
  {https://ui.adsabs.harvard.edu/abs/2015ApJ...801...33T} {801, 33}

\bibitem[\protect\citeauthoryear{{Torii} et~al.,}{{Torii}
  et~al.}{2017}]{2017ApJ...835..142T}
{Torii} K.,  et~al., 2017, \mn@doi [\apj] {10.3847/1538-4357/835/2/142}, \href
  {https://ui.adsabs.harvard.edu/abs/2017ApJ...835..142T} {835, 142}

\bibitem[\protect\citeauthoryear{{Truelove}, {Klein}, {McKee}, {Holliman},
  {Howell}, {Greenough}  \& {Woods}}{{Truelove}
  et~al.}{1998}]{TrueloveEtAl1998}
{Truelove} J.~K.,  {Klein} R.~I.,  {McKee} C.~F.,  {Holliman} John~H. I.,
  {Howell} L.~H.,  {Greenough} J.~A.,   {Woods} D.~T.,  1998, \mn@doi [\apj]
  {10.1086/305329}, \href
  {https://ui.adsabs.harvard.edu/abs/1998ApJ...495..821T} {495, 821}

\bibitem[\protect\citeauthoryear{{Tubbs}}{{Tubbs}}{1982}]{Tubbs1982}
{Tubbs} A.~D.,  1982, \mn@doi [\apj] {10.1086/159846}, \href
  {https://ui.adsabs.harvard.edu/abs/1982ApJ...255..458T} {255, 458}

\bibitem[\protect\citeauthoryear{{Turk}, {Smith}, {Oishi}, {Skory}, {Skillman},
  {Abel}  \& {Norman}}{{Turk} et~al.}{2011}]{TurkEtAl2011}
{Turk} M.~J.,  {Smith} B.~D.,  {Oishi} J.~S.,  {Skory} S.,  {Skillman} S.~W.,
  {Abel} T.,   {Norman} M.~L.,  2011, \mn@doi [\apjs]
  {10.1088/0067-0049/192/1/9}, \href
  {https://ui.adsabs.harvard.edu/abs/2011ApJS..192....9T} {192, 9}

\makeatother
\end{thebibliography}




\appendix

\section{Cloud properties: comparison with observations}
\label{sec:Cloud properties: comparison with observations}

\begin{figure*}
	\includegraphics[width=\columnwidth]{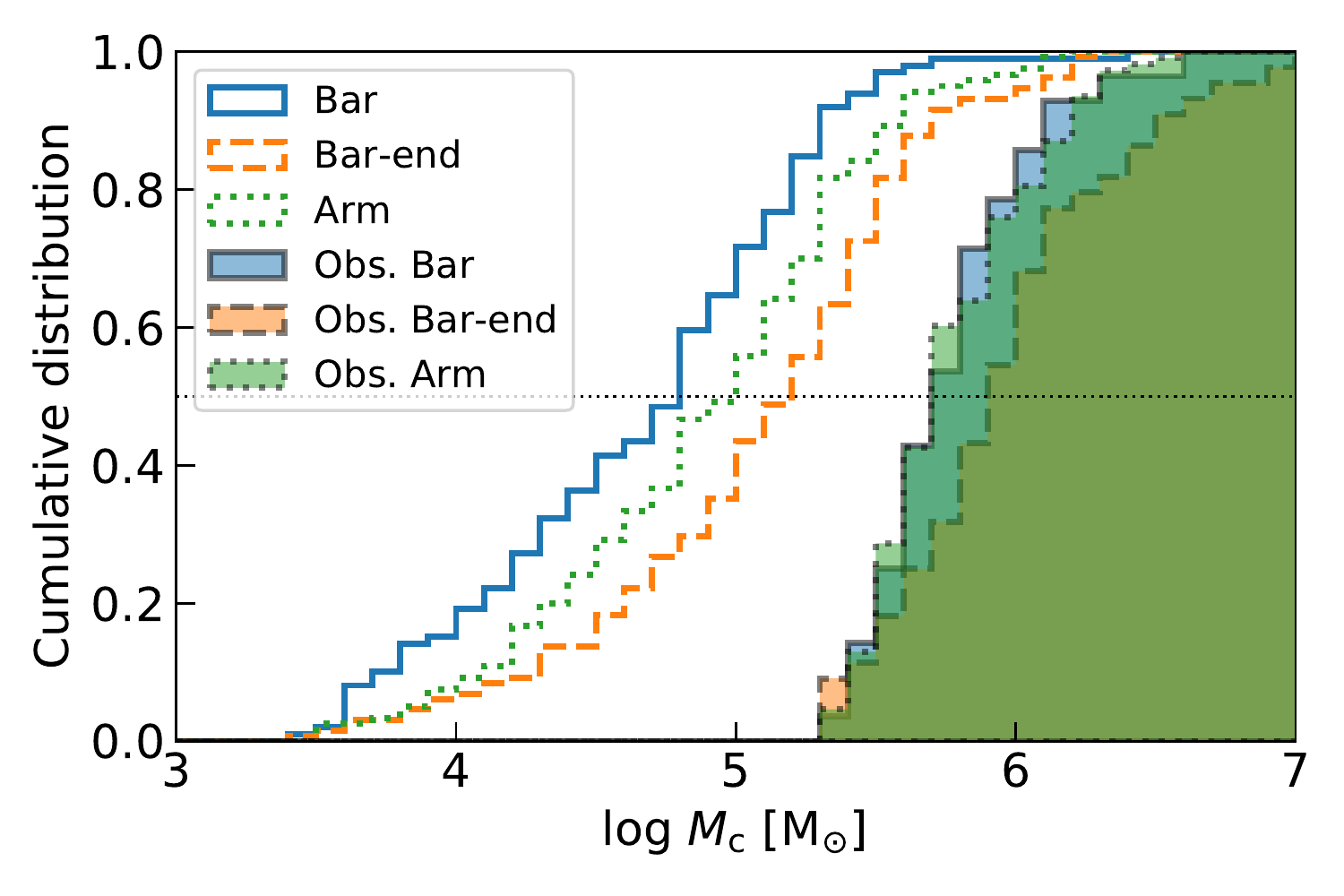}
	\includegraphics[width=\columnwidth]{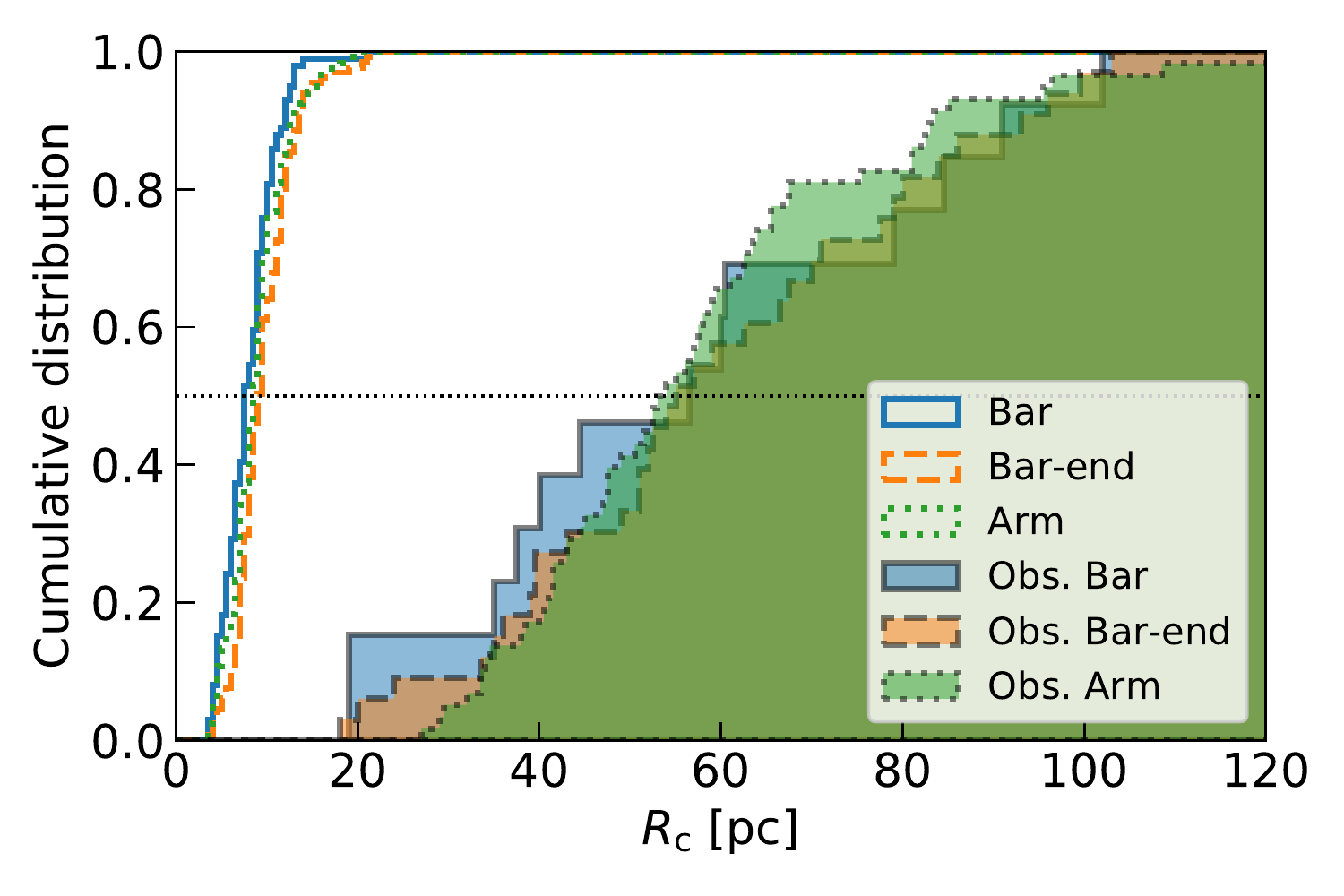}
	\includegraphics[width=\columnwidth]{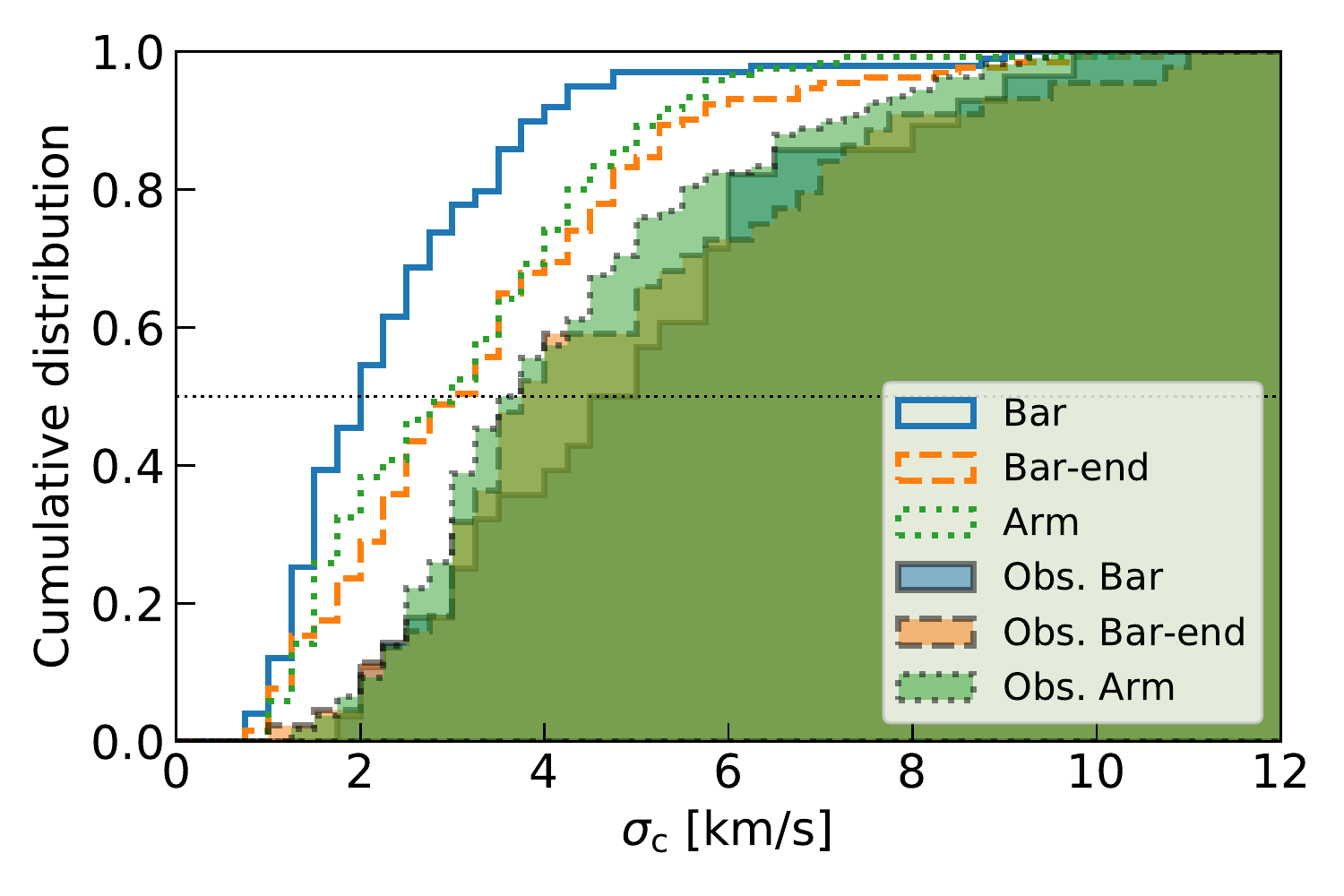}
	\includegraphics[width=\columnwidth]{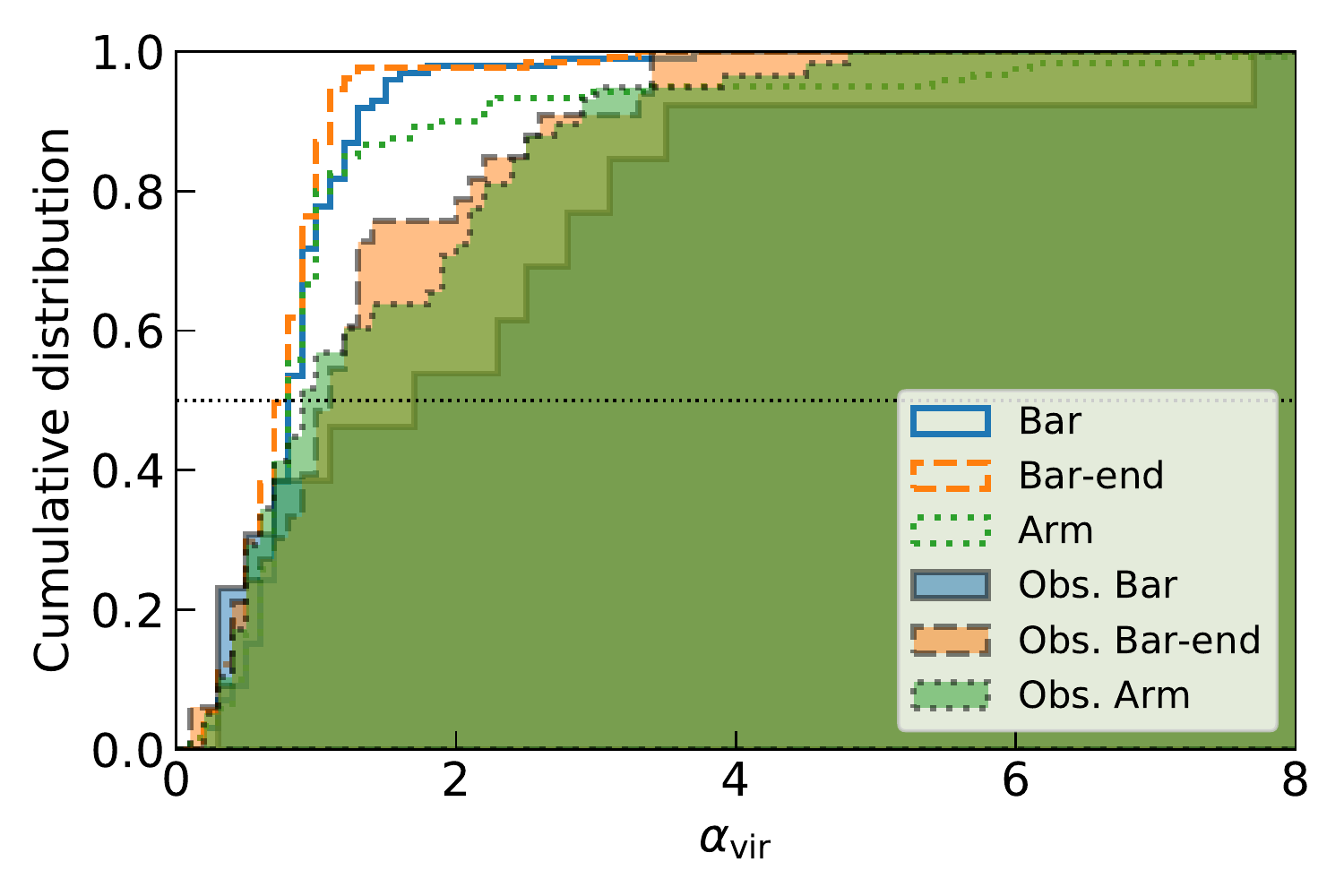}
    \caption{Normalized cumulative distribution function of cloud masses (top left), radii (top right), velocity dispersion (bottom left), and virial parameters (bottom right), with the observed clouds in the same galaxy of NGC1300 observed by \citet{MaedaEtAl2020}. Note that the beam size is 40 pc, and the mass completed limit is $2.0 \times 10^5\ \mathrm{M_{\odot}}$ for the observed samples.}
    \label{fig:cloud_properties_compare_obs_1}
\end{figure*}

\begin{figure*}
	\includegraphics[width=\columnwidth]{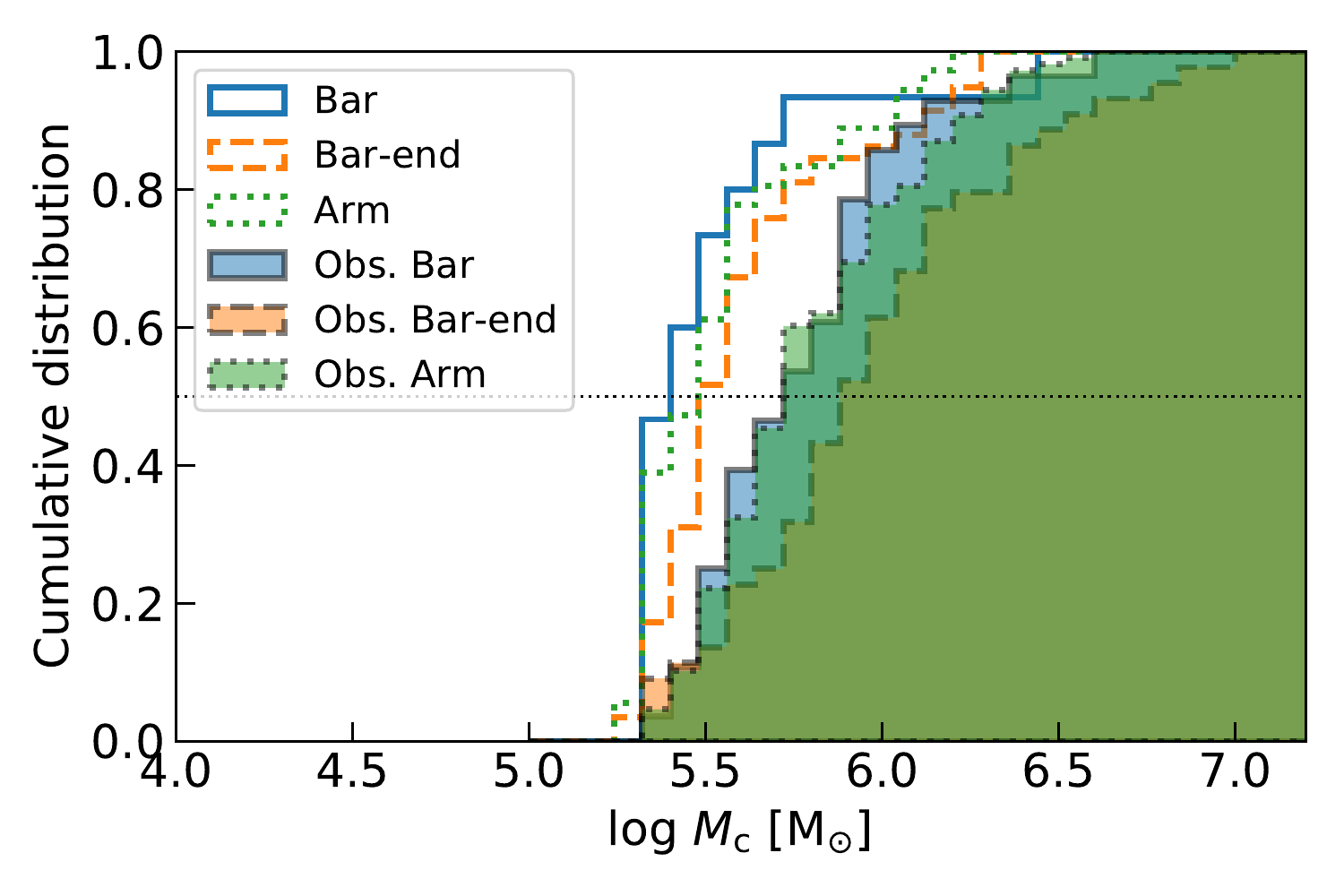}
	\includegraphics[width=\columnwidth]{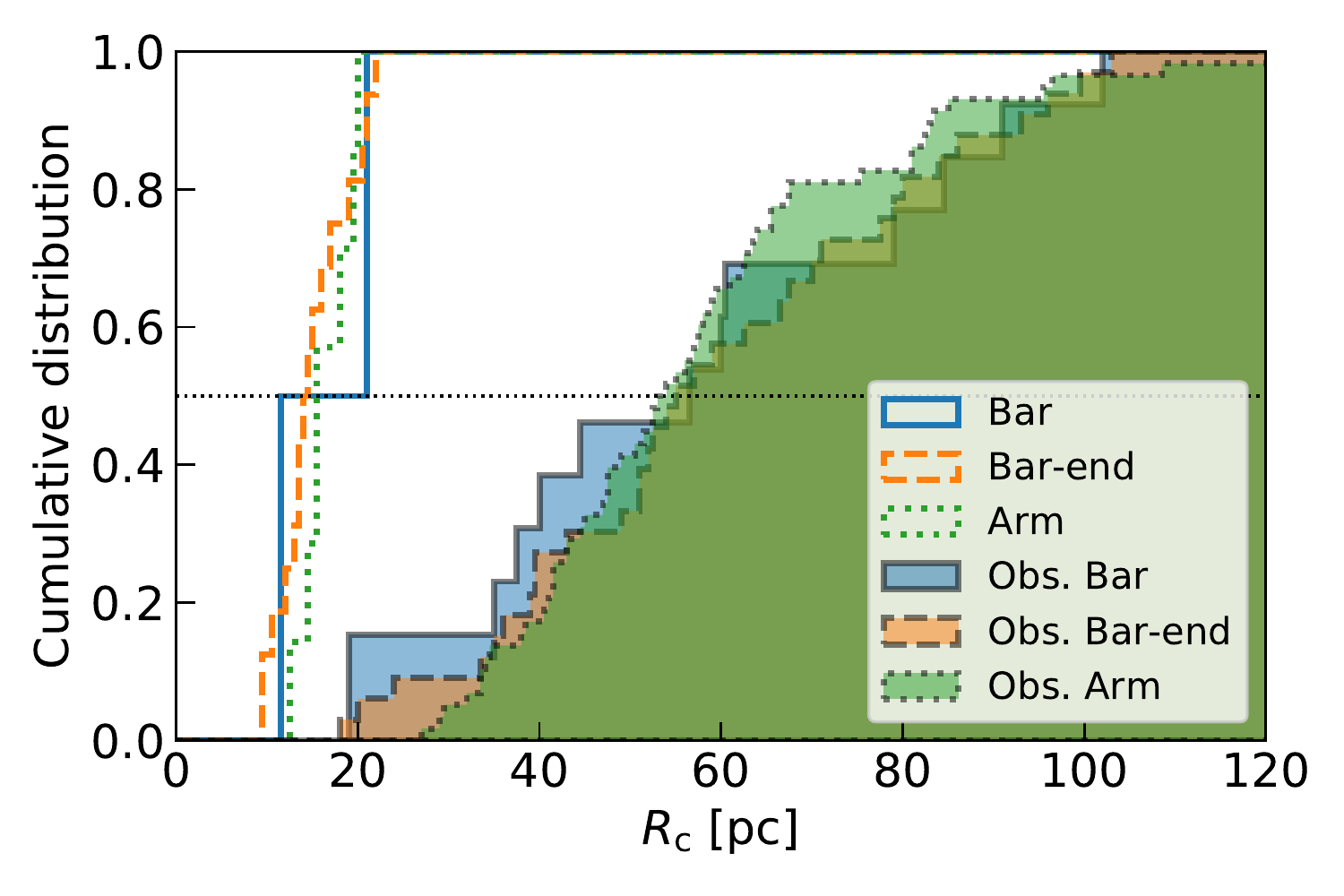}
	\includegraphics[width=\columnwidth]{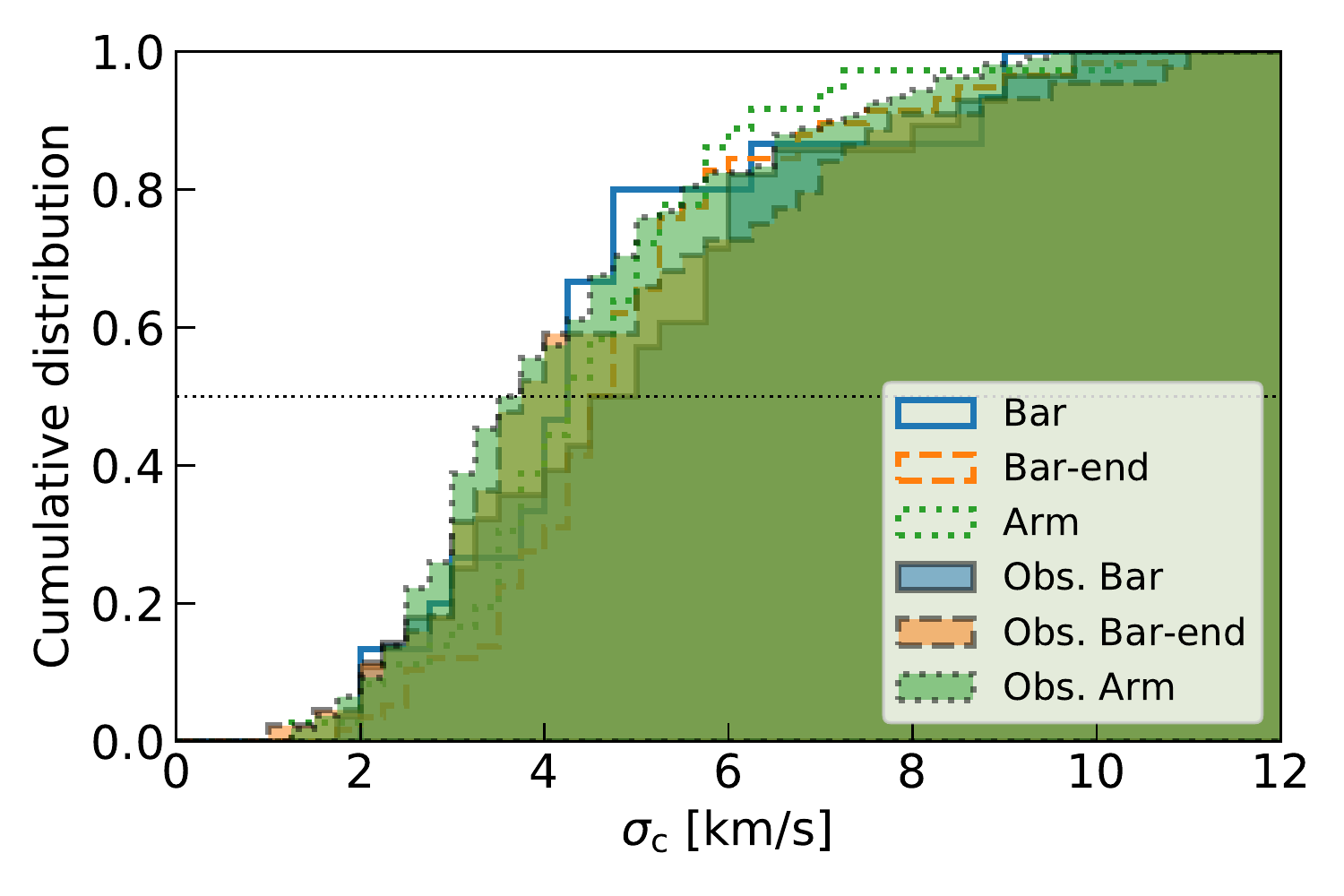}
	\includegraphics[width=\columnwidth]{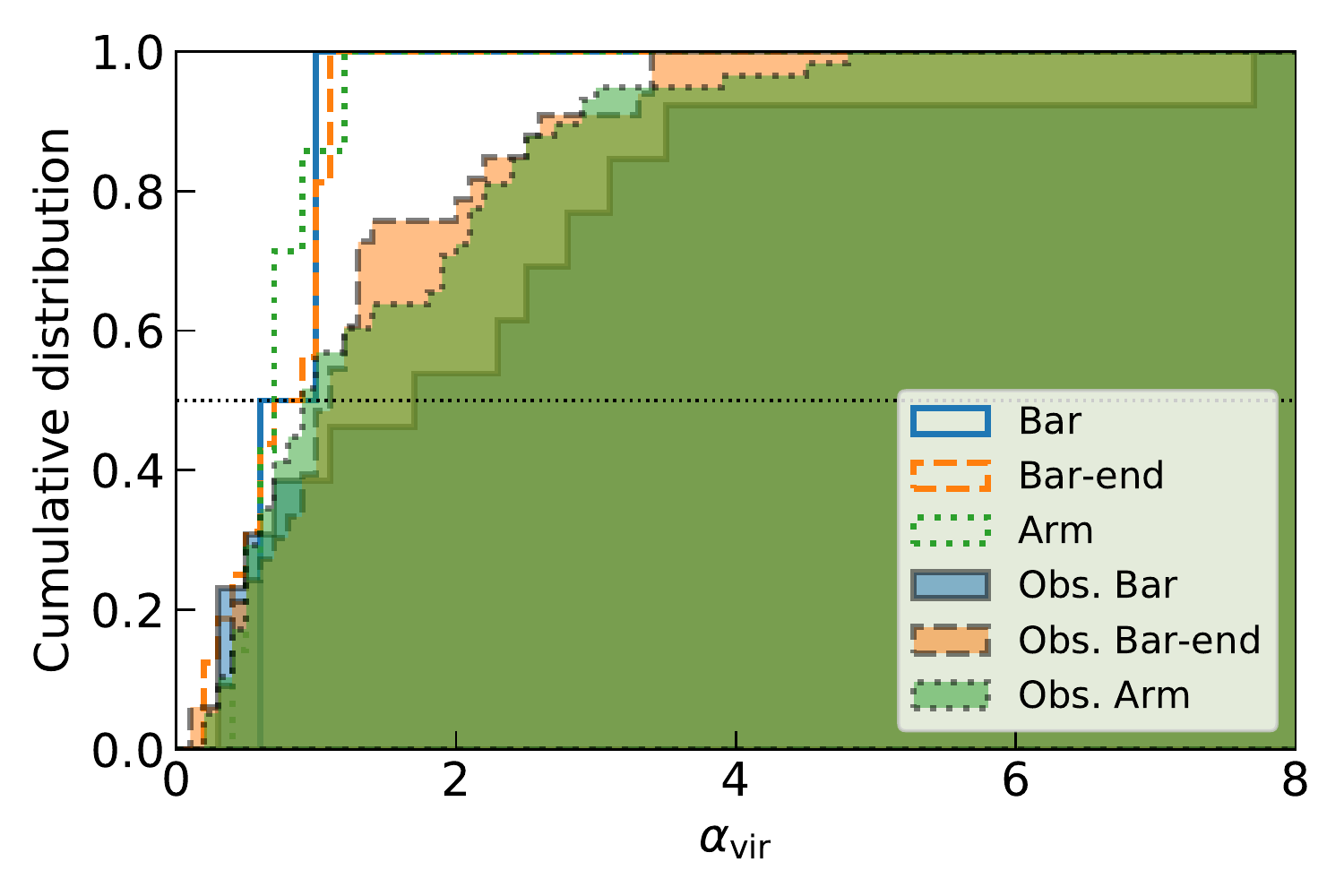}
    \caption{Same as Figure~\ref{fig:cloud_properties_compare_obs_1}, but showing the simulated clouds with $M_{\mathrm{c}} > 2.0 \times 10^5\ \mathrm{M_{\odot}}$ for cloud masses and velocity dispersion, and with $M_{\mathrm{c}} > 5.0 \times 10^5\ \mathrm{M_{\odot}}$ for cloud radii and virial parameters, which are the same selection criteria used for the observed clouds.}
    \label{fig:cloud_properties_compare_obs_2}
\end{figure*}

We compare the cloud properties with the observational results of the same galaxy of NGC1300 done by \citet{MaedaEtAl2020}. They had done $^{12}$CO($1-0$) observations toward the western bar, arm and bar-end regions with ALMA 12-m array with a beam size of $\sim$ 40 pc. With identification of GMCs using the CPROPS algorithm, they found that there was virtually no significant variations in GMC properties among the galactic regions. 

Figure~\ref{fig:cloud_properties_compare_obs_1} shows direct comparisons between simulated and observed clouds. Over all, the simulation and observation have the same qualitative features as regards environmental dependence of cloud properties; there is no significant systematic difference between three galactic regions. However, looking at the mean values, there is quantitative difference between simulation and observation. The cloud masses and radii in the simulation are lower than those in the observation; the difference can be one order of magnitude for mass and at least a factor of six for radius.

The situation gets better if we use the same selection criteria used for the observed clouds. In the observation, the mass completeness limit was estimated to be $2.0 \times 10^5\ \mathrm{M_{\odot}}$, and the mass limit for resolved samples was $5.0 \times 10^5\ \mathrm{M_{\odot}}$ \citep[see Section 5 in][]{MaedaEtAl2020}. Therefore, the observed cloud samples were selected with $M_{\mathrm{c}} > 2.0 \times 10^5\ \mathrm{M_{\odot}}$ for masses and velocity dispersion, and with $M_{\mathrm{c}} > 5.0 \times 10^5\ \mathrm{M_{\odot}}$ for radii and virial parameters. Figure~\ref{fig:cloud_properties_compare_obs_2} shows the same normalized cumulative functions of cloud properties, but the simulated clouds are selected with the same selection criteria which are used for the observed clouds. The mean masses for the simulated clouds get close to those for the observed clouds, although the simulated clouds are still smaller and less massive.

Having a systematic difference between simulation and observation is not surprising because the spatial resolution of the ALMA observation is at most 40 pc, which is much larger than the 2 pc resolution of this simulation. The low resolution in the observation could cause non-detection of small clouds or merging smaller clouds to a connected larger structure. We should also mention that the simulated clouds can be relatively small due to the slightly high threshold density of $n_{\mathrm{H, c}} = 400\ \mathrm{cm}^{-3}$ for cloud definition. For those reasons, the simulation and observation might be looking at somewhat different objects; it is possible that some observed clouds could be associations which consist of diffuse ambient molecular gas and/or multiple clouds, and, on the other hand, the simulated clouds could be each dense clumps of which the association consists. For a more complete comparison, there are many things we should do; e.g. we need to convolve the simulation with the beam size of the observation and fine-tune the threshold density for the cloud definition, as mentioned above. We should also consider the method for cloud identification, which is currently different between observation and simulation: the cloud in the simulation is identified as a coherent structure contained within a contour at the threshold density no matter how many density peaks exist within the cloud, but the observed cloud splits up into multiple clouds when there are multiple density maxima. Moreover, post-processing of the simulation data for a chemical transition from atomic gas to molecular gas or radiative transfer to calculate the CO emission might be needed as well. 

With regards to internal velocity dispersion, the difference between simulation and observation is minimal, in particular, in the case we impose the selection criteria of $M_{\mathrm{c}} > 2.0 \times 10^5\ \mathrm{M_{\odot}}$, as shown in Figure~\ref{fig:cloud_properties_compare_obs_2}. That might be because we calculate a mass-weighted velocity dispersion for simulated clouds, so that the ambient low density gas surrounding the denser parts of the clouds hardly affects the over-all value of the velocity dispersion. As a results, we get the virial parameter for the simulated clouds close to those for the observation, as shown in both Figure~\ref{fig:cloud_properties_compare_obs_1} and Figure~\ref{fig:cloud_properties_compare_obs_2}.


\section{Example of fast collision}
\label{sec:Example of fast collision}

\begin{figure*}
	\includegraphics[width=300pt]{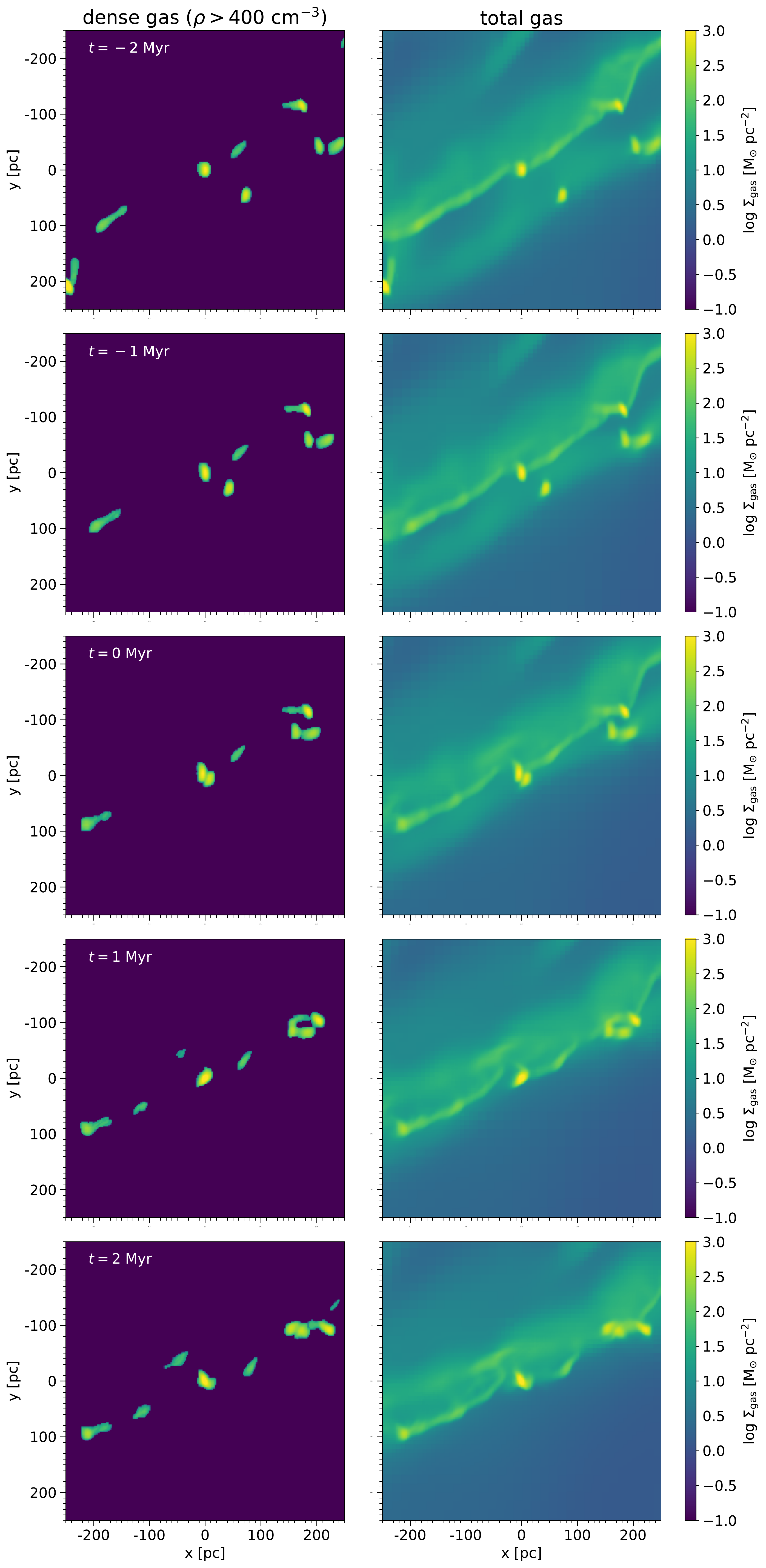}
    \caption{One example of the fast cloud collision occurred in the \textit{Bar} region. Top to bottom, the images show the time evolution of the collision. Left panels show the dense cloud gas ($\rho > 400\ \mathrm{cm^{-3}}$), and the right panels show the total gas.}
    \label{fig:collision_projections}
\end{figure*}

Figure~\ref{fig:collision_projections} shows one example of the fast collision occurred in the \textit{Bar} region at $t = 598.8$ Myr. Although most gentle collisions in the \textit{Bar-end} and \textit{Arm} regions occur in filaments, some fast collisions in the \textit{Bar} region are triggered by filament-filament collisions, which is the case shown in this figure. The collision speed is 34.5 km/s, and masses of the colliding clouds are $2.66 \times 10^5\ \mathrm{M_{\odot}}$ and $9.55 \times 10^4\ \mathrm{M_{\odot}}$. The massive cloud has the internal velocity dispersion of 3.8 km/s and a virial parameter of 0.9 just before collision at $t = -2$ or $-1$ Myr. During the collision ($t = 0$ Myr), the velocity dispersion and virial parameter increase to 9.6 km/s and 4.3, respectively. After the collision ($t = 2$ Myr), they get back to normal: 4.7 km/s and 0.9.


\bsp	
\label{lastpage}
\end{document}